\documentclass[epj]{svjour}

\usepackage{amsfonts, amssymb, amsmath, graphicx, comment, bm, slashed, caption, subcaption, dcolumn,color}
\usepackage{graphics}
\newcommand{\beq}{\begin{eqnarray}}
\newcommand{\eeq}{\end{eqnarray}}

\newcommand{\bd}{\mathbf}

\newcommand{\calP}{ {\cal P}}

\newcommand{\lqcd}{\Lambda_{ {\rm QCD}} }
\newcommand{\Nc}{N_{ {\rm c}} }
\newcommand{\Nf}{N_{ {\rm f} } }
\newcommand{\rmd}{ {\rm d} }
\newcommand{\rmi}{ {\rm i} }
\newcommand{\rme}{ {\rm e} }
\newcommand{\la}{\langle}
\newcommand{\ra}{\rangle}
\newcommand{\calH}{ {\cal H} }
\begin{document}
\title{Phenomenological neutron star equations of state}
\subtitle{3-window modeling of QCD matter}
\author{Toru Kojo 
}                     
%
%
\institute{Department of Physics, University of Illinois at Urbana-Champaign, 1110 W. Green Street, Urbana, Illinois 61801, USA}
\date{Received: date / Revised version: date}
%
\abstract{We discuss the 3-window modeling of cold, dense QCD matter equations of state at density relevant to neutron star properties. At low baryon density, $n_B \lesssim 2n_s$($n_s$: nuclear saturation density), we utilize purely hadronic equations of state that are constrained by empirical observations at density $n_B\sim n_s$ and neutron star radii. At high density, $n_B \gtrsim 5n_s$, we use the percolated quark matter equations of state which must be very stiff to pass the two-solar mass constraints. The intermediate domain at $2 \lesssim n_B/n_s \lesssim 5$ is described as neither purely hadronic nor percolated quark matter, and the equations of state are inferred by interpolating hadronic and percolated quark matter equations of state. Possible forms of the interpolation are severely restricted by the condition on the (square of) speed of sound, $0\le c_s^2 \le 1$. The characteristics of the 3-window equation of state are compared with those of conventional hybrid and self-bound quark matters. Using a schematic quark model for the percolated domain, it is argued that the two-solar mass constraint requires the model parameters to be as large as their vacuum values, indicating that the gluon dynamics remains strongly non-perturbative to $n_B\sim 10n_s$. The hyperon puzzle is also briefly discussed in light of quark descriptions.
\PACS{
      {PACS-key}{discribing text of that key}   \and
      {PACS-key}{discribing text of that key}
     } 
} 
\maketitle
\section{Introduction}
\label{intro}

Neutron stars are the cosmic laboratories for the studies of quantum chromodynamics (QCD) at low temperature and high density \cite{Lattimer:2006xb,Buballa:2014jta,Fukushima:2010bq}. Solving Tolmann-Oppenheimer-Volkoff (TOV) equation together with the QCD equations of state, we can calculate the neutron star mass-radius ($M$-$R$) relation which can be confronted with the observations. At given conditions such as rotation, magnetic field, temperature, each equation of state gives the unique $M$-$R$ relation. This procedure is invertible; using Lindblom's algorithm \cite{Lindblom1992}, we can extract an equation of state uniquely from a given $M$-$R$ curve \cite{Chen:2015zpa}.
%

For conventional neutron stars, the shapes of $M$-$R$ curves are strongly correlated with pressures at several fiducial densities \cite{Lattimer:2000nx,Ozel:2009da}. Pressure around $n_B \sim 1-2n_s$ ($n_B$: baryon density\footnote{Notations: ($n_B$, $\mu_B, \cdots$) with subscript $B$ are used for baryonic quantities, and ($n$, $\mu, \cdots$) for quarks.},  $n_s\simeq 0.16\,{\rm fm}^{-3}$: nuclear saturation density) determines the overall size of typical neutron stars, and beyond $\sim 2n_s$ the $M$-$R$ curve moves toward the vertical direction without large variation of the radius. Its slope is correlated with pressure at $n_B \sim 2-4n_s$. The curve eventually ends at the maximal mass which is primarily determined by pressure at $n_B \gtrsim 4n_s$. Thus the slope and maximum mass carry most direct information for dense matter beyond the conventional nuclear matter domain.
%
\begin{figure*}
\hspace{2.50cm}
\includegraphics[width = 0.7\textwidth]{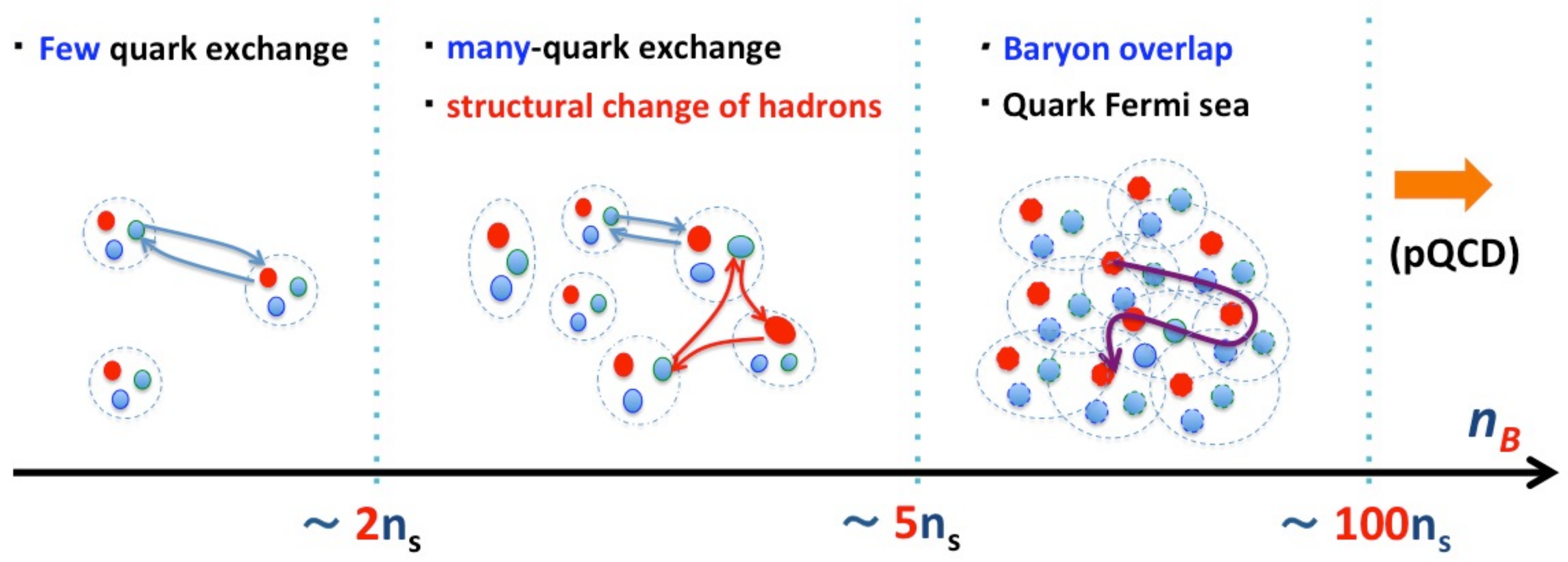}
\vspace{-0.1cm}
\caption{
\footnotesize{The 3-window description for QCD matter. (i) At $n_B \lesssim 2n_s$, nucleons are well separated and exchange only few quarks (or mesons). (ii) At $n_B \gtrsim 2n_s$, baryons other than nucleons start to appear and baryons exchange many quarks, developing many-body forces. Accordingly there are structural changes in hadronic wavefunctions. (iii) At $n_B \gtrsim 5 n_s$ the baryons begin to percolate and the quark Fermi sea develops. Until $n_B$ reaches $\sim 100n_s$, the matter remains in the strongly correlated regime.
}
\vspace{-0.1cm}
}\label{fig:3-window}
\end{figure*}

In this respect, the discoveries of two-solar mass ($2M_\odot$) neutron stars \cite{2m_1,2m_2} have significant impacts. The existence of such stars mean that the QCD equations of state at $n_B>2n_s$ are very stiff, i.e., the pressure ($P$) at given energy density ($\varepsilon$) is very large to resist the gravitational collapse. In fact a variety of soft high density equations of state have been ruled out by this mass constraint alone. The neutron star radii should put further constraint \cite{Steiner:2010fz,Steiner:2012xt,Ozel:2010fw,Ozel:2015fia,Guillot:2014lla,Heinke:2014xaa,Suleimanov:2010th}, although the determination contains more uncertainties than the mass measurement.

There are several purely nucleonic equations of state that can pass the $2M_\odot$ constraint. But typical hadronic models, in their typical versions, indicate that hyperons are likely to appear at $n_B>2-3n_s$, and they greatly soften the equations of state; in general the massive degrees of freedom, near its mass threshold, adds mass times its number density to the energy density, but adds only little amount of pressure. The conflict between the $2M_\odot$ constraint and hyperon softening is called "hyperon puzzle" \cite{Massot:2012pf,Schulze:2011zza,Weissenborn:2011kb,Lonardoni:2014bwa,Nishizaki:2001in}.
%

Similar softening problems occurs for typical hybrid equations of state, in which hadronic and weakly interacting quark matter are treated as distinct and the full equation of state is made through the Maxwell construction for the first order phase transition. Typical models predict the phase transition occur at $n_B \sim 2-10 n_s$. Since the first order transition has a jump in energy density at fixed pressure, it always softens the equations of state. This calls debates on the existence of quark matter at the neutron star core \cite{Ozel:2006bv,Alford:2006vz}.
%

We shall take up these problems as great opportunities to study a matter which is neither purely hadronic nor weakly interacting quark matter. To be clear, we emphasize that the purpose of this paper is {\it not} to predict the equations of state. Rather, our primary purpose is to delineate the properties of strongly correlated QCD matter and develop a plausible picture for non-perturbative QCD at finite density. This will be done by examining {\it supposed} equations of state that are being more and more constrained by the neutron star observations and the QCD predictions from the chiral effective theory (ChEFT) and perturbative QCD (pQCD). Then we will argue which aspects of the QCD matter are relevant to resolve the aforementioned softening problems.
%

To infer the properties of such matter, we first note that the softening problems occur somewhat outside of the domain of applicability of the conventional descriptions. In this respect, we should refer to the chiral effective theory \cite{Weinberg:1978kz,Epelbaum:2008ga,Bogner:2009bt,Gezerlis:2014zia,Gandolfi:2015jma,Gandolfi:2011xu} and pQCD \cite{Fraga:2001id,Freedman:1976ub,Freedman:1977gz,Kurkela:2009gj}. These methods can provide not only quantitative predictions but also predict the domain of applicability; in particular these methods can address when the underlying pictures behind the calculations require modifications. The hyperon problem occurs around $n_B \sim 2-3n_s$ within purely hadronic descriptions. But such descriptions, at $n_B>2n_s$, may cause problems, or at least raise questions, on the convergence of many-body forces, appearance of new baryonic degrees of freedom, and possible structural changes in the QCD vacuum and hadronic wavefunctions. To understand all these issues, it is important to go back to more microscopic description from which effective theories are derived. On the other hand, the quark matter problem occurs at $n_B \sim 2-10 n_s$ within the description of weakly interacting quark matter. But at such density the weakly interacting picture for quark matter is not quite valid as can be seen from pQCD equations of state; the current state-of-art calculations to $O(\alpha_s^2)$ clarify that the interactions are important already at relatively large quark chemical potential $\mu \sim 1\,{\rm GeV}$ or $n_B \sim 10^2 n_s$. Moreover the pQCD at $n_B \lesssim 10^2 n_s$ results start to show fairly large renormalization scale dependence\footnote{Or the convergence problems at $\mu \lesssim 1\,{\rm GeV}$ can be clearly seen from the studies of Ref.\cite{Fraga:2001id} where the authors compared the $O(\alpha_s)$ and $O(\alpha_s^2)$ corrections.}, indicating that in quark descriptions the matter at $n_B \simeq 2-10n_0$ should be regarded as strongly correlated matter.
%

These observations bring us to the 3-window description for the non-perturbative domain of QCD matter\footnote{More modest interpolation approach is to connect hadronic equations of state to pQCD as worked out in Refs.\cite{Freedman:1977gz,Kurkela:2009gj,Kurkela:2014vha}. While these studies have more predictive characters, our primary concern is how to interpret such pressure curves. These two approaches are complementary.}, which was introduced by Masuda, Hatsuda, and Takatsuka \cite{Masuda:2012kf}, and was also discussed in Ref. \cite{Alvarez-Castillo:2013spa,Hell:2014xva,Kojo:2014rca} in slightly different versions. Below we will follow the descriptions of Ref.\cite{Kojo:2014rca}, whose schematic picture is summarized in Fig.\ref{fig:3-window}. We use purely hadronic equations of state at low density where baryons exchange only few mesons (or quarks). To be specific, at $n_B<2n_s$, we use the Akmal-Pandharipande-Ravenhall (APR) equation of state, which is purely nucleonic \cite{Akmal:1998cf}. At density high enough for the overlap of baryons or percolation\footnote{If we take the volume of a baryon to be $V=4\pi r_s^3/3$ with the radius $r_s$, the baryons overlap at $n_B \simeq 3-12n_s$ for $r_s \simeq 0.5-0.8\,{\rm fm}$, assuming them to be static objects. The percolation would occur at lower density because their momenta are not quite small.} \cite{Baym:1976yu}, we construct the equations of state using effective quark models at the QCD scale $\lqcd\simeq 0.2\, {\rm GeV}$. To be specific, at $n_B>5n_s$, we use the 3-flavor Nambu-Jona-Lasinio (NJL) model plus vector and diquark interactions. Consequently we are left with the domain of $2n_s<n_B<5n_s$ for which appropriate pictures are not known even at qualitative level. Such matter is supposedly described as hadronic matter about to percolate, or quark matter about to be confined. We infer the equations of state by interpolating the APR and NJL pressures as functions of chemical potential. In this work we will use smooth interpolation assuming the hadron-quark crossover \cite{Schafer:1998ef,Kitazawa:2002bc,Hatsuda:2006ps,Zhang:2008wx}, although the 3-window construction itself does not reject the possibility of the first order phase transition as far as the jump of the energy density is not too large. For simplicity, we use a polynomial form for the interpolated pressure and require it to smoothly match with the APR and NJL pressures up to the second order of derivatives.
%

In addition to the boundary conditions, the interpolated equations of state must obey  several physical conditions. The interpolated pressure must (i) be stiff enough to allow the existence of $2M_\odot $ neutron stars; (ii) yield the square of speed of sound, $c_s^2 = \partial P/\partial \varepsilon$, which is positive and smaller than the square of speed of light, $0\le c_s^2\le 1$. The reasonable interpolation satisfying these conditions is possible only for a restricted class of quark matter equations of state at high density. We will study what kind of interactions are relevant for such quark matter equations of state, and will show that such interactions follow from the knowledge of hadron spectroscopy and nuclear physics, provided that the gluon dynamics at finite quark density is not significantly different from that in the QCD vacuum \cite{McLerran:2007qj,Kojo:2011fh}.
%

This paper is structuralized as follows. In Sec.\ref{sec:pre} we begin with preliminary remarks. We present graphical analyses of $P(\mu)$ curves which allow us to get rough but quick insights on the stiffness and speed of sound for given equations of state. This method appears to be very useful to classify several types of equations of state. In Sec.\ref{sec:class}, we examine the characteristics of the 3-window equations of state and argue the quark matter softening problem. We discuss why the 3-window construction can give a stiff high density equation of state, by allowing quark matter equations of state stiffer than those typically appearing in the conventional hybrid construction. The observation is especially important when we consider soft hadronic equations of state at low density. The self-bound quark matter equations of state are also discussed for further comparisons. 
In Sec.\ref{sec:APR} we briefly explain a model of nuclear and crust equations of state. In Sec.\ref{sec:model} we introduce a schematic quark model for percolated quark matter. The model consists of the standard NJL model plus additional vector and diquark interactions for baryon and nuclear phenomenology. We examine the impacts of the additional interactions and show that these contributions are important to express the sensible equations of state. In Sec.\ref{sec:3-window} we construct 3-window equations of state and compute the $M$-$R$ relations by solving TOV-equation. We will show how severely the conditions of speed of sound constrain interpolated equations of state and thereby the range of model parameters. Sec.\ref{sec:discussions} is devoted to conjectures and interpretations of our equations of state. We discuss the hyperon problem from the viewpoint of quark descriptions. Physics behind our stiff quark matter equations of state is also discussed in the context of gluon dynamics at finite density. The Sec.\ref{sec:summary} is devoted to summary and outlook.

%

\section{Preliminary remarks}
\label{sec:pre}

Following Ref.\cite{Kojo:2014rca}, we consider equations of state in the grand canonical ensemble and work with $P(\mu)$. In this choice of thermodynamic variables, the problems 
in quark matter softening will become most manifest. The thermodynamic ground state has the maximal pressure at given $\mu$. If there are several candidates of phases, we can just plot them in the plane of $P$ v.s. $\mu$ and can easily identify the phase transition point. On the other hand, the discussions of the stiffness ($P$ v.s. $\varepsilon$) require a little more efforts; we need to calculate $n(\mu)=\partial P/\partial \mu$ and then use the relation $\varepsilon(\mu) = \mu n -P$ to extract $\varepsilon$. 
%

In this section we discuss a graphical method which can be used to get quick estimate of stiffness from a $P(\mu)$ curve. The method is also useful to get rough insights for the conditions of the speed of sound. Thus using a $P(\mu)$ curve alone, we can see the relation among phase transitions, stiffness, speed of sound, and then can understand how these conditions put constraints one another. 
%

\subsection{Stiffness from $P(\mu)$ curve}
\label{stiff}

%
\begin{figure}
\hspace{1.2cm}
\resizebox{0.35\textwidth}{!}{%
  \includegraphics{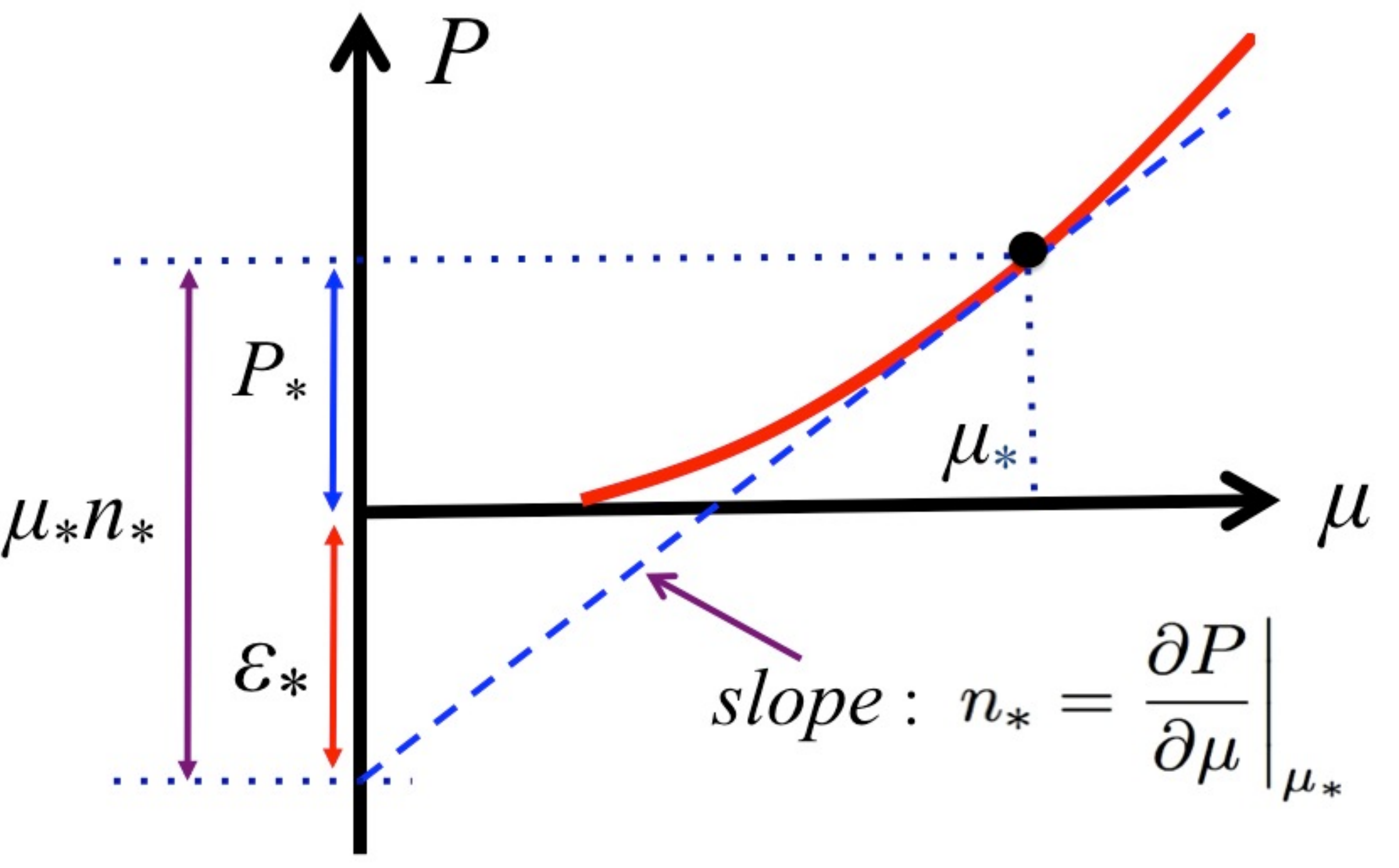}
}
\caption{The energy density $\varepsilon$ from $P(\mu)$ curve.}
\label{fig:p_mu_e}       
\end{figure}

Suppose some $P(\mu)$ curve. Our task is to estimate $\varepsilon$ at some value of chemical potential, $\mu_*$. To find out $\varepsilon_*=\varepsilon(\mu_*)$, we notice that the tangential line of the $P(\mu)$ curve at $\mu=\mu_*$, with the slope $n_*=\partial P/\partial \mu|_{\mu=\mu_*}$,  intercepts with the $P$-axis at $\varepsilon_*$. This can be easily seen from Fig.\ref{fig:p_mu_e}.
%

%
\begin{figure}
\hspace{1.2cm}
\resizebox{0.35\textwidth}{!}{%
  \includegraphics{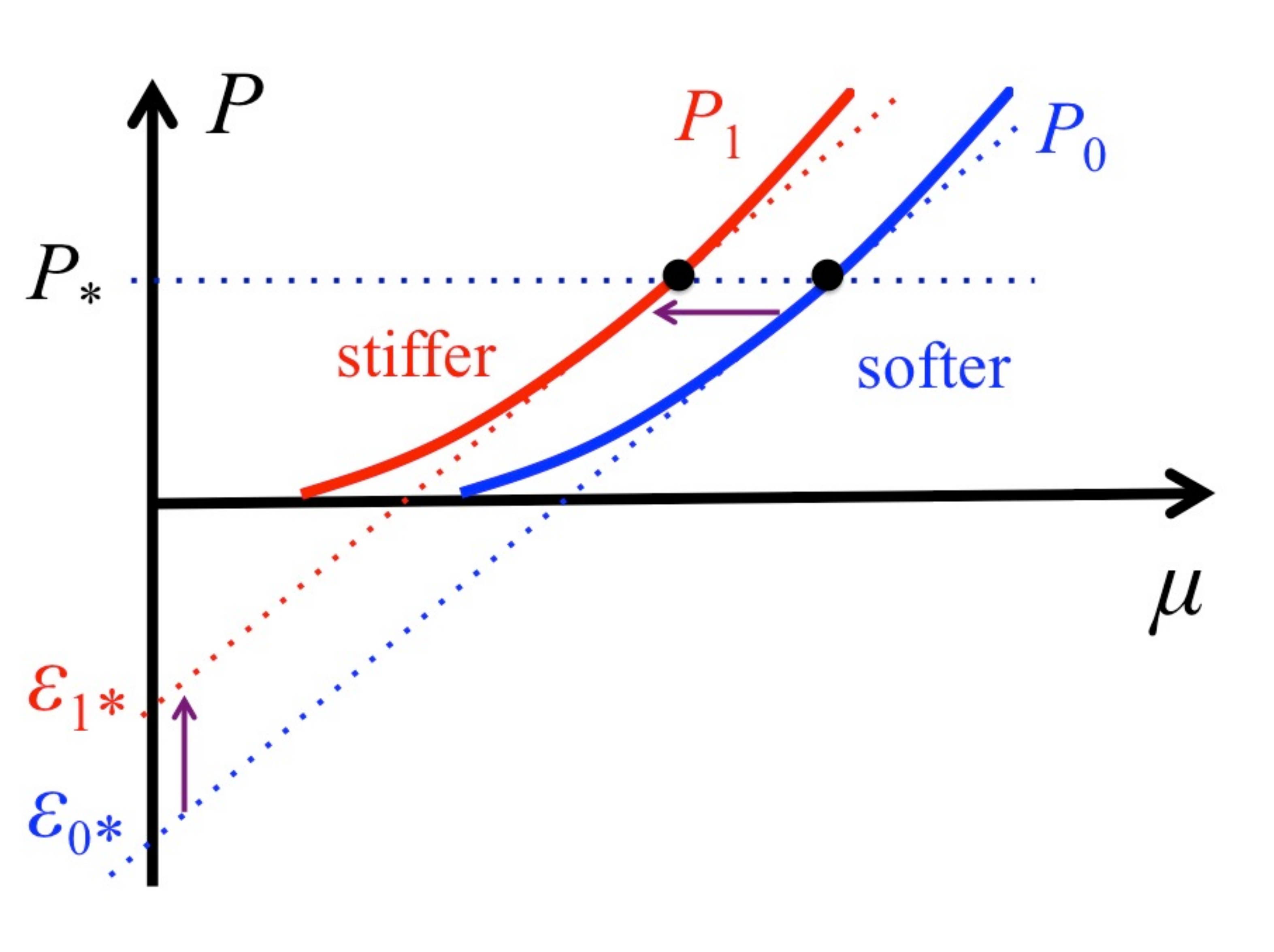}
}
\caption{Stiff ($P_1$) v.s. soft ($P_0$) equations of state. At given pressure $P_*$, the curve $P_1$ gives less energy density than the $P_0$ curve.}
\label{fig:p_mu_e_eg1}       
\end{figure}

To see the utility, let us apply the method to several examples. The first example is the bag model with the bag constant $B (>0)$. This constant must be added to (subtracted from) energy density (pressure) when the ground state becomes perturbative, because the perturbative vacuum has larger energy than the non-petrubative vacuum. The pressure in the perturbative domain is given by $P = c_0 \mu^4 - B$, and from which $n=4c_0\mu^3$ and $\varepsilon = 3c_0\mu^4+B$ follow. Since $P=(\varepsilon-4B)/3$, larger $B$ leads to softer equations of state. In the graphical method, this conclusion follows immediately as seen from Fig.\ref{fig:p_mu_e_eg1} ; at fixed value of $P_*=P(\mu_*)$, a curve $P_0$, which has a larger bag constant $B$ than $P_1$, leads to the larger length for the intercept for the tangential line. This means that $P_1$ is stiffer than $P_0$.
%

%
\begin{figure}
\hspace{1.2cm}
\resizebox{0.35\textwidth}{!}{%
  \includegraphics{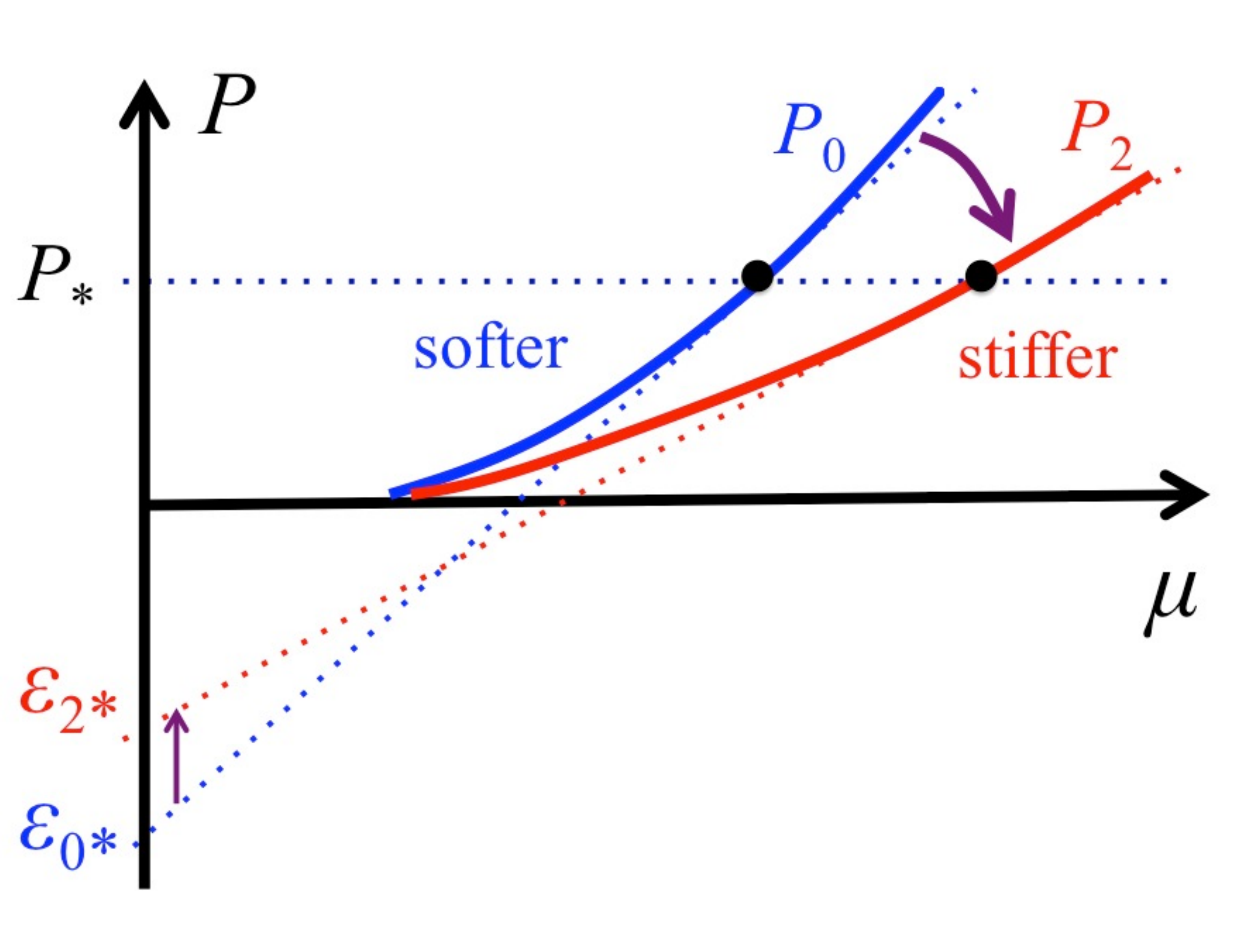}
}
\caption{Stiff ($P_2$) v.s. soft ($P_0$) equations of state. $P_2$ is a clockwise rotated curve of $P_0$.}
\label{fig:p_mu_e_eg2}       
\end{figure}

For another example, suppose a curve $P_2$ which is obtained by rotating $P_0$ clockwise around the point where $P_0=0$ (Fig.\ref{fig:p_mu_e_eg2}). Clearly, the length of the intercept of the tangential line gets smaller after the rotation, meaning that $P_2$ is stiffer than $P_0$. This sort of behaviors occurs when we include repulsive density-density interactions; more chemical potential (or energy density) is required to increase the number density and pressure. The resulting $P(\mu)$ curve grows slowly as $\mu$ increases.
%
 
Later we will examine the impacts of various interactions on $P(\mu)$ curves. The results may be understood using the above two examples.
%

\subsection{Physical speed of sound from $P(\mu)$}
\label{causality}

%
\begin{figure}
\hspace{1.2cm}
\resizebox{0.35\textwidth}{!}{%
  \includegraphics{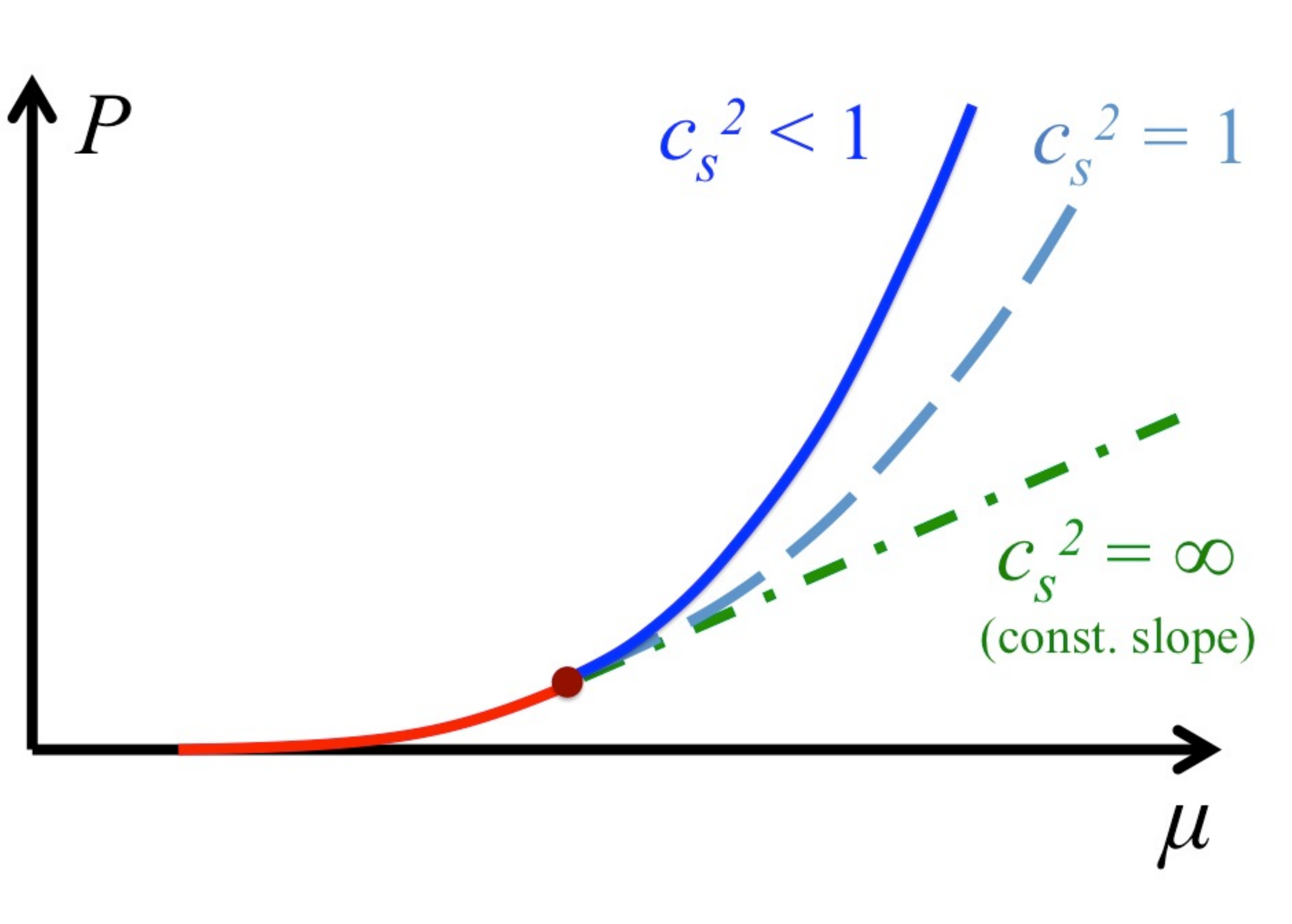}
}
\caption{The schematic plot for $P(\mu)$ curves with fixed sound velocities. At some chemical potential (indicated by the dot), we switch from some curve to fixed sound velocity curves. For equations of state to be causal, the $P(\mu)$ curves must grow faster than quadratic powers.}
\label{fig:p_mu_cs2_eg1}       
\end{figure}

Equations of state are under the constraints from the general relativity and thermodynamics; the square of speed of sound $c_s^2 = \partial P/\partial \varepsilon $ must be positive and less than the square of the speed of light, $0 \le c_s^2 \le 1$. We examine how to see these constraints in the $P(\mu)$ curves.
%

To do so, first we notice the thermodynamic relations, $\rmd P = n \rmd \mu$ and $\rmd \varepsilon = \mu \rmd n$, from which one gets
\beq
c_s^2 = \frac{\, \partial P \,}{\partial \varepsilon} = \frac{\, \partial \ln \mu \,}{\partial \ln n} \,.
\eeq
Suppose a small region around $(n_*, \mu_*)$ where $c_s$ is approximately constant. In such a region, we can solve the differential equation to get
\beq
n = n_* \left( \frac{\mu}{\mu_*} \right)^{1/c_s^2}\,.
\eeq
For instance, $n\propto \mu^3$ for an ideal gas with $c_s^2=1/3$, or $n\propto \mu$ for the causal limit $c_s^2=1$, or $n={\rm const.}$ for infinite speed of sound, $c_s^2 = \infty$.
Note also that if $n$ is a decreasing function of $\mu$, the square of the speed of sound is negative and unphysical, $c_s^2<0$. 
%

In this example, the pressure at $\mu\simeq \mu_*$ is ($P_*=P(\mu_*)$)
\beq
P \simeq P_* + {\rm const.} \times \mu^{1/c_s^2 + 1} \,.
\eeq
For an equation of state to be causal ($c_s^2\le 1$), the pressure must grow faster than $\mu^2$ at any points on the $P(\mu)$ curve (Fig.\ref{fig:p_mu_cs2_eg1}).
%

%
\begin{figure}
\hspace{1.2cm}
\resizebox{0.35\textwidth}{!}{%
  \includegraphics{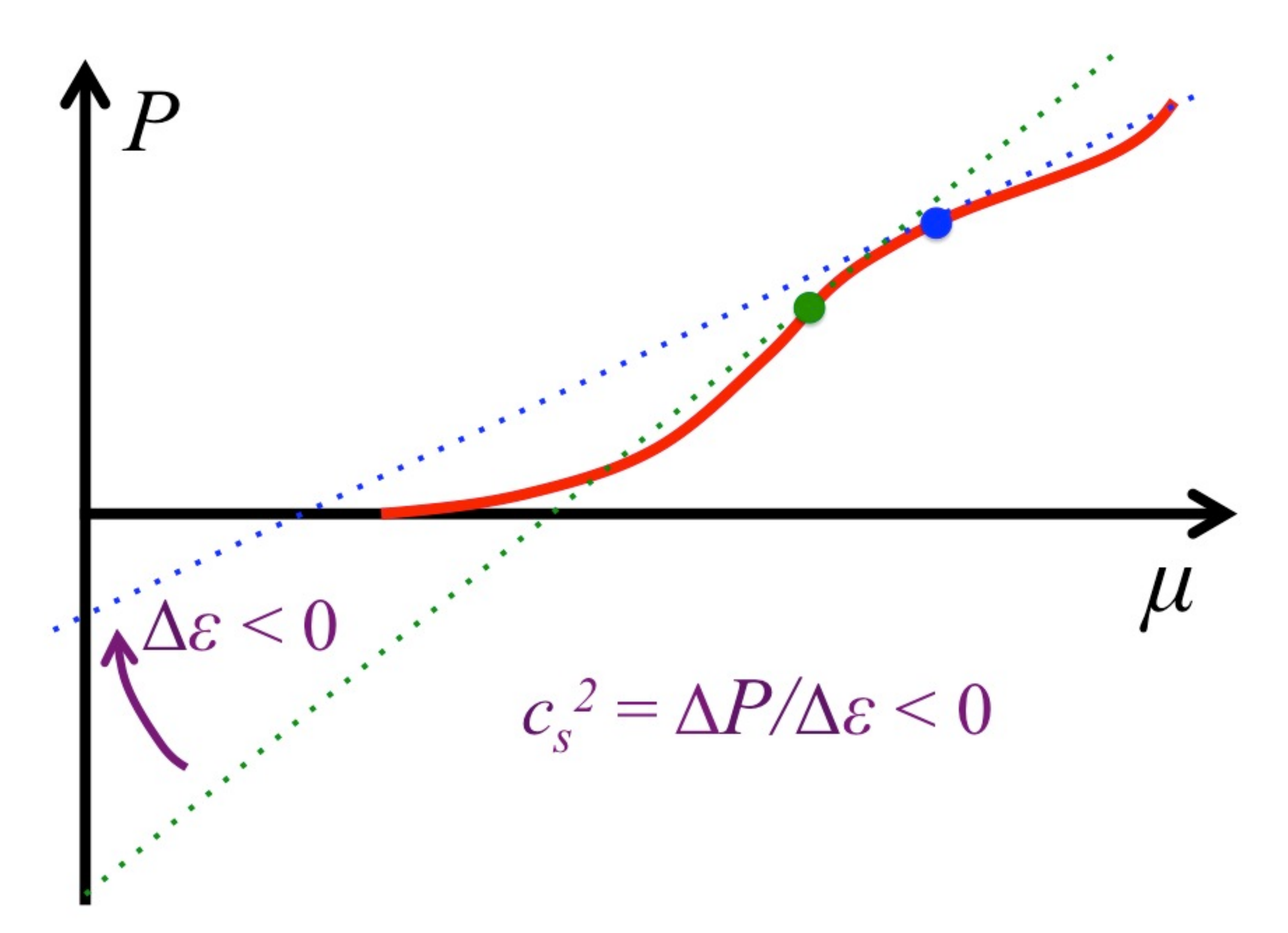}
}
\caption{A unphysical $P(\mu)$ with an inflection point. The region with $\partial^2 P/\partial \mu^2 <0$ has the unphysical speed of sound, $c_s^2<0$. }
\label{fig:p_mu_cs2_eg2}       
\end{figure}

In Ref.\cite{Kojo:2014rca}, the authors imposed the positive susceptibility constraint, $\partial^2 P/\partial \mu^2 = \partial n/\partial \mu >0$. This is equivalent to requiring the speed of sound to be positive, because
\beq
\frac{\, \partial^2 P \,}{\partial \mu^2} = \frac{\, \partial n\,}{\partial \mu}  =  \frac{\, {\rm const.} \,}{c_s^2} \times \mu^{1/c_s^2 - 1} \,,
\eeq
becomes negative only if $c_s^2<0$. The positive susceptibility condition can be easily examined by checking whether $P(\mu)$ curves contain inflection points or not\footnote{To see this, in most plots we will not divide $P(\mu)$ by $\mu^4$ or the ideal gas pressure.} (Fig.\ref{fig:p_mu_cs2_eg2}).
%

As we will see, the interpolation in the 3-window construction is tightly constrained by these conditions. In turn, the quark matter equations of state, which supply the boundary condition for the interpolated curves, are also severely restricted. 
%

\section{Classification of quark equations of state}
\label{sec:class}

In this section we classify the equations of state that contain quark matter at the neutron star core, and review how they pass the $2M_\odot$ constraint in different ways. The conventional hybrid, 3-window, and self-bound quark matter equations of state are discussed. 
%

\subsection{The conventional hybrid equations of state}
\label{con_eos}

%
\begin{figure}
\hspace{1.2cm}
\resizebox{0.35\textwidth}{!}{%
  \includegraphics{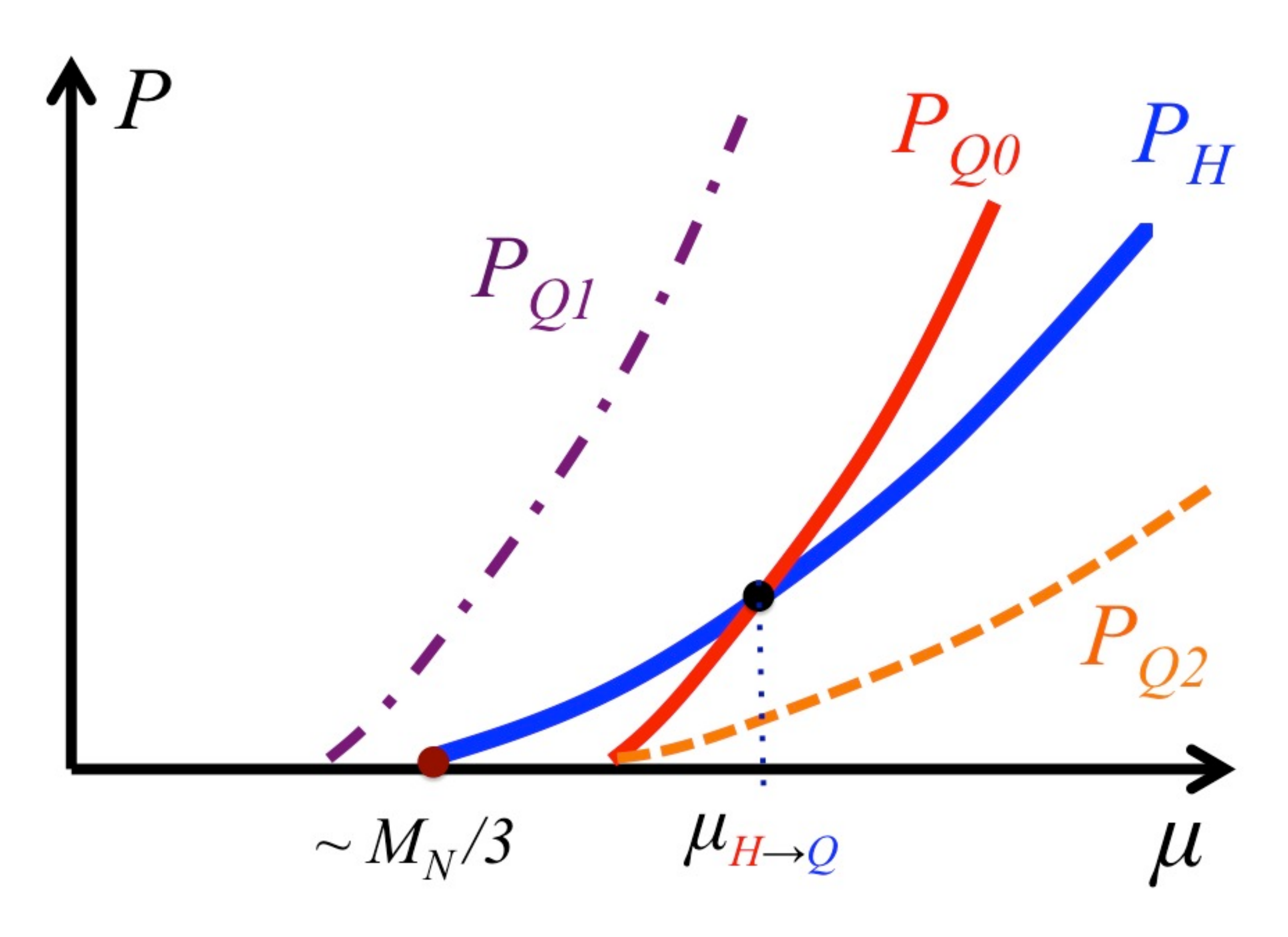}
}
\caption{A hadronic pressure $P_H$ and quark equations of state, $P_{Q0}$, $P_{Q1}$, and $P_{Q_2}$. In the conventional hybrid construction, only $P_{Q0}$ can be used; the other curves are stiff (see Figs.\ref{fig:p_mu_e_eg1} and \ref{fig:p_mu_e_eg2}), but cannot be combined with $P_H$.  }
\label{fig:p_mu_hybrid}       
\end{figure}

In the conventional construction of hybrid equations of state, one uses hadronic and quark matter equations of state as distinct. At some density the matter undergoes the first order phase transition from hadronic to quark matter phase. To describe such situation, we must use a quark matter equation of state such that its pressure $P(\mu)$ approaches the hadronic one from below; quark pressure must grow sufficiently fast, otherwise we would not get the intersection point (Fig.\ref{fig:p_mu_hybrid}). As we saw in the last section, such quark equations of state are typically soft, having troubles with the $2M_\odot$ constraint. This tendency has called debates on whether quark matter can exist at the core of neutron stars \cite{Ozel:2006bv,Alford:2006vz}.
%

%
\begin{figure}
\hspace{1.2cm}
\resizebox{0.35\textwidth}{!}{%
  \includegraphics{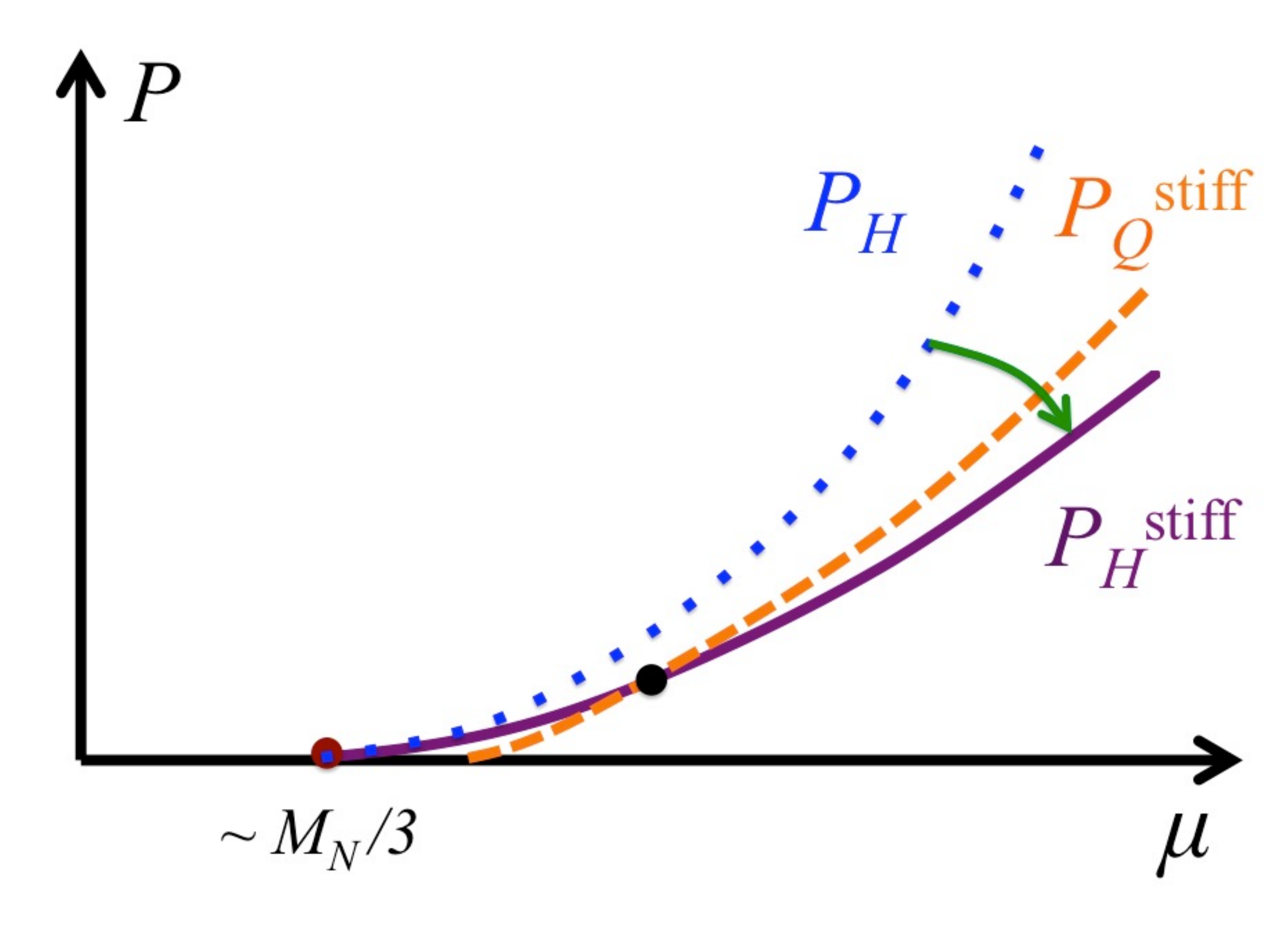}
}
\caption{An example of stiff hybrid equations of state. Assuming the hadronic equation of state is stiffened at high density, stiff quark equation of state can find the intersection point at low density.}
\label{fig:p_mu_hybrid_eg2}       
\end{figure}
\begin{figure}
\hspace{1.3cm}
\resizebox{0.35\textwidth}{!}{%
  \includegraphics{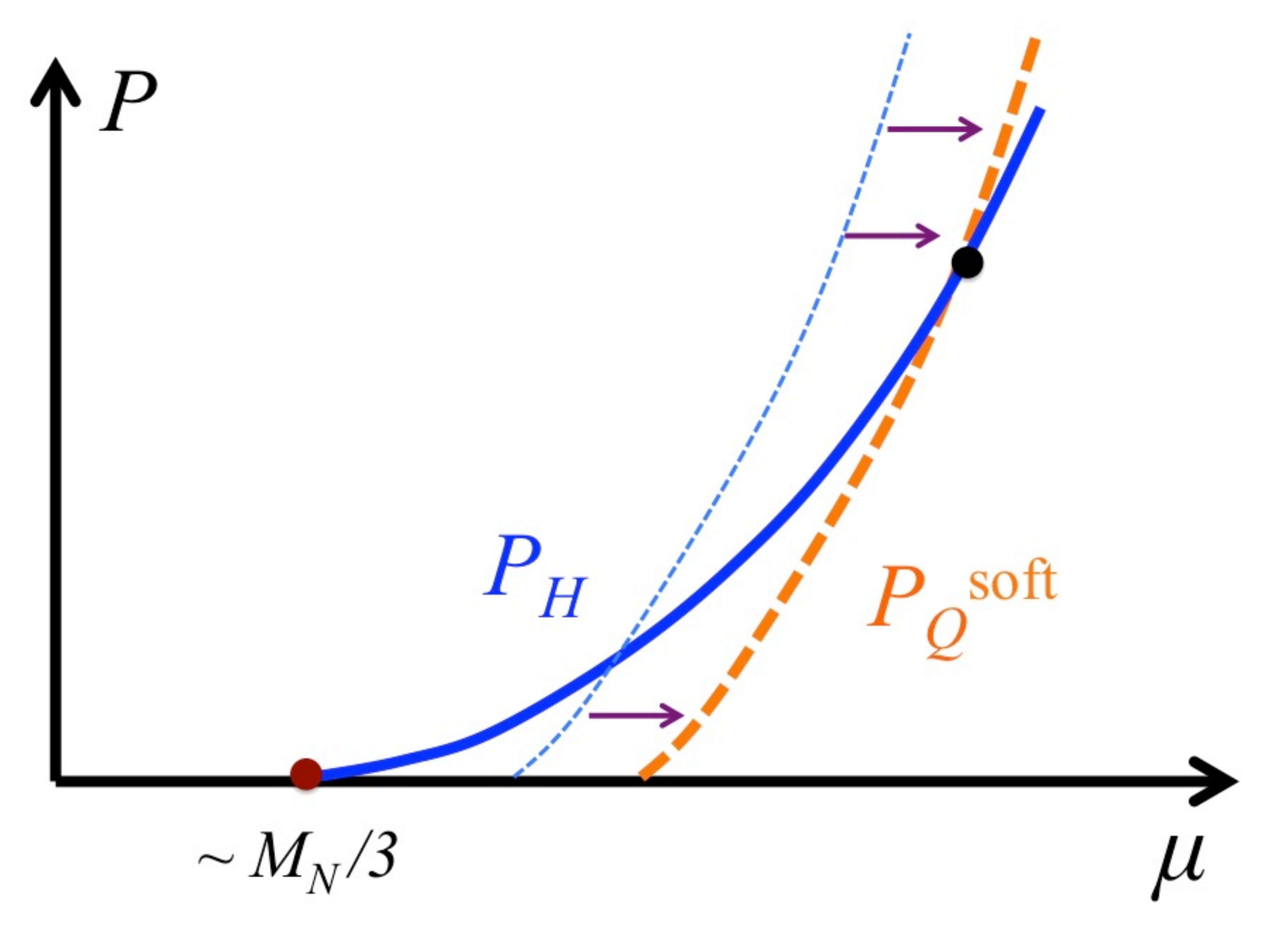}
}
\caption{An example of stiff hybrid equations of state. Assuming the intersection of $P_H$ and $P_Q$ occur at very high density, hadronic equations of state, which are assumed to be stiff at very high density (like purely nucleonic equations of state),  build up most of the neutron star mass before the quark matter softening occurs.}
\label{fig:p_mu_hybrid_eg3}       
\end{figure}

There are several ways to bypass this stiffness problem. For instance, we can allow stiff quark equations of state if hadronic equations of state extrapolated to high density are stiff \cite{Klahn:2006iw,Bonanno:2011ch,Benic:2014jia}. In that case the hadronic pressure grows so slowly that even a stiff quark pressure curve (which does not grow very fast) can have the intersection with the hadronic one (Fig.\ref{fig:p_mu_hybrid_eg2}). In this stiff-stiff combination, it is possible to pass the $2M_\odot$ constraint with large amount of quark core (although some cares are necessary for the causality condition). Another way to pass the $2M_\odot$ constraint is to use stiff high density hadronic equations of state and to locate the first order phase transition point at very high density \cite{Akmal:1998cf}. In this case, the stiff hadronic matter builds most of the neutron star mass before the quark matter softening occurs (Fig.\ref{fig:p_mu_hybrid_eg3}). In this case the amount of quark matter core is tiny.
%

While the above treatments can satisfy the $2M_\odot$ constraint, there is fundamental weakness in the conventional hybrid construction. The problem is that we must compare two equations of state even though their domains of applicability might not overlap; purely hadronic picture is valid at low density while at high density its applicability is questionable; quark matter picture is valid at high density but at low density it would be affected by large non-perturbative corrections such as confining effects. It is potentially dangerous to use high density hadronic equations of state to select out acceptable quark matter equations of state. Also one cannot rule out quark matter equations of state even if their low density behaviors are at odd with nuclear phenomenology; the origin of problem might be just the wrong extrapolation of correct high density quark equations of state. 
%

Finally, let us note that the conventional construction implicitly omits an equation of state which is neither purely hadronic nor quark matter. The 3-window construction, which will be discussed next, will deal with this omitted possibility.
%

\subsection{The 3-window equations of state}
\label{3window_eos}

%
\begin{figure}
\hspace{1.2cm}
\resizebox{0.35\textwidth}{!}{%
  \includegraphics{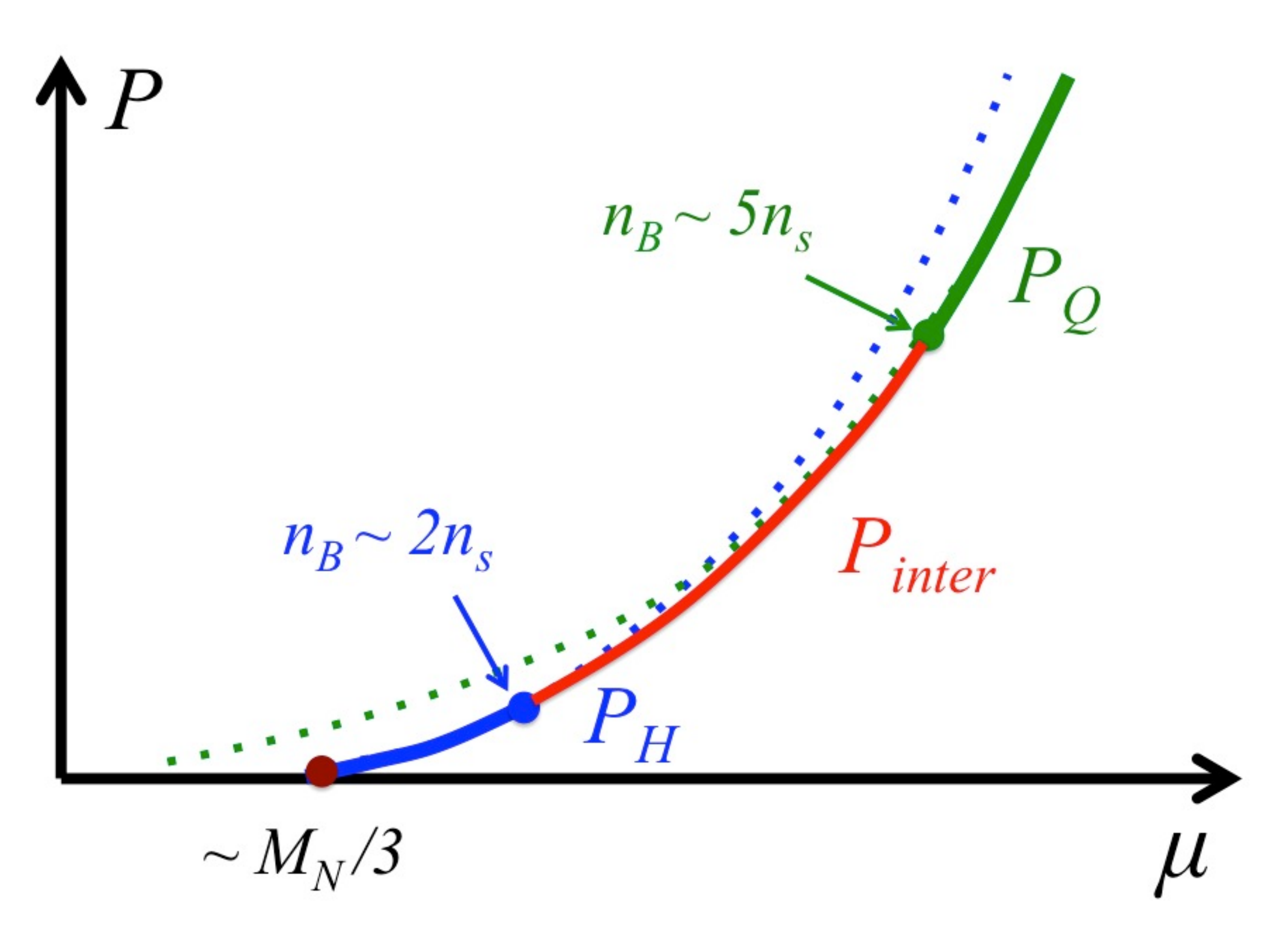}
}
\caption{An example of 3-window equations of state. The boundaries are given at $n_B\sim 2n_s$ and $n_B \sim 5n_s$ for hadronic and percolated quark pressure, respectively. The 3-window construction accepts pressure curves forbidden in the conventional construction, and allows us to explore a new class of (stiff) quark matter equations of state.}
\label{fig:p_mu_inter}       
\end{figure}

The 3-window construction is based on the picture that (at $n_B<10^2 n_s$) QCD matter may consist of three domains; purely hadronic matter at low density, percolated quark matter at high density, and matter intermediate between these two matters.
%

In practice, at low density we use equations of state purely based on the hadronic degrees of freedom. Hadrons are proper effective degrees of freedom to describe the system as far as the system is dilute enough. Baryons are well-separated and exchange only few mesons (quarks), so that the interactions are tractable and modifications of hadronic wavefunctions are negligible. Also there are only few excited baryonic states appearing in the virtual processes, and they can be manifestly taken into account or can be integrated out into a few numbers of effective interactions. These treatments, however, are difficult to justify at high density. In fact there are no established hadronic equations of state beyond $\sim 2n_s$. Thus we will stop using the purely hadronic descriptions for density above $n_B \simeq 2n_s$.
%

At high density, we use equations of state based on the percolated quark matter picture. Baryons overlap and quarks no longer belong to specific baryons. Then quarks develop the Fermi sea. The percolated matter picture, however, cannot be justified at density smaller than $n_B \sim 5n_s$; three quarks gather to form a baryon. The confining effects become crucial so we will stop using the percolated quark matter descriptions for density below $\sim 5 n_s$.
%

There is an intermediate region for $2 n_s \lesssim  n_B  \lesssim 5 n_s$ where neither of the above pictures are applicable. The matter may be described as strongly interacting hadronic matter, or hadronic matter about to percolate, or quark matter about to be confined. To infer the corresponding equations of state, we simply interpolate the equations of state of purely hadronic and percolated quark matters.
%

In the 3-window construction, we have more chances to get stiff quark matter equations of state than in the conventional hybrid one (Fig.\ref{fig:p_mu_inter}). For instance, the {\it extrapolated} hadronic equations of state at high density will not be used to rule out some of stiff quark matter equations of state. Also we will not judge the validity of a quark equation of state by its low density behavior because it should acquire huge corrections at low density.  Any quark equations of state are acceptable as far as their behavior at high density is appropriate and provide the proper boundary conditions that allow the interpolated equations of state to be physical. These observations enable us to study a new class of quark equations of state which have not been fully explored, and the resulting quark equations of state can be stiff.
%

%
\begin{figure}
\hspace{1.2cm}
\resizebox{0.33\textwidth}{!}{%
  \includegraphics{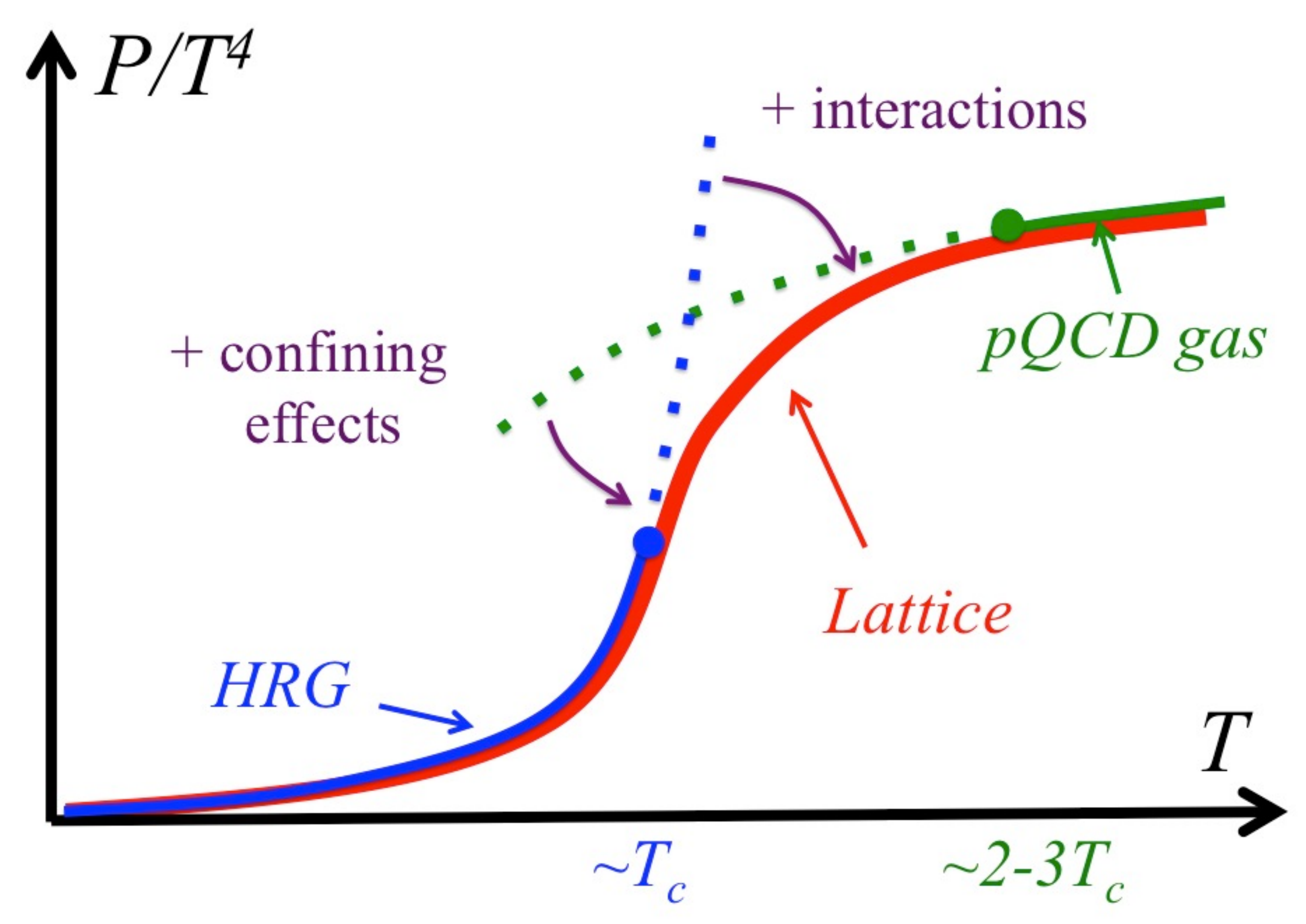}
}
\caption{The schematic plot for the equation of state in finite temperature QCD. By interpolating the hadron resonance gas (HRG) and (resummed) pQCD pressure, the 3-window pressure gives good qualitative description for the lattice data.}
\label{fig:inter_finiteT}  
\end{figure}

In order to make the 3-window arguments more convincing, here we quote the finite temperature QCD equation of state to use some analogy (Fig.\ref{fig:inter_finiteT}. See also Refs.\cite{Asakawa:1995zu,Fukushima:2014pha}). At low temperature, the hadron resonance gas (HRG) model, which is a gas of non-interacting hadrons, gives very good description for the pressure of the lattice QCD data \cite{Huovinen:2009yb,Ding:2015ona,Bazavov:2012jq,Bazavov:2013dta}. Around the (pseudo-)critical temperature $T_c$ for chiral (or deconfinement) transition, however, the HRG pressure starts to overshoot the lattice data. This is because around $T_c$ thermally excited hadrons overlap and the interactions are not negligible; the inclusion of the interactions tempers the overshooting behavior. On the other hand, at high temperature the (resummed)pQCD theories give reasonable descriptions for the lattice data \cite{Blaizot:2000fc,Andersen:2011sf}. Below $2-3T_c$, however, the pQCD pressure tends to overshoot the lattice data\footnote{The overshooting behavior is clearly seen in pure Yang-Mills for $T\le 2-3T_c^{ {\rm YM} } \simeq 0.5-0.8\, {\rm GeV}$, while with quarks this tendency depends on the choice of the renormalization scale; the hard thermal loop results for LO, NLO pressure, and NNLO result with relatively small $\alpha_s$ show the overshooting behaviors, but NNLO with large $\alpha_s$ gives the undershooting behaviors \cite{Andersen:2011sf}. But the latter has large deviation from NLO pressure, and its validity is a bit uncertain.}. This is not surprising, because the pQCD calculations do not include the confining effects which trap quarks into hadrons. The inclusion of the confining effects such as the Polyakov loop can tame the problem by suppressing artificial excess of the pQCD pressure \cite{Fukushima:2003fw}.
%

%
\begin{figure}
\hspace{1.2cm}
\resizebox{0.33\textwidth}{!}{%
  \includegraphics{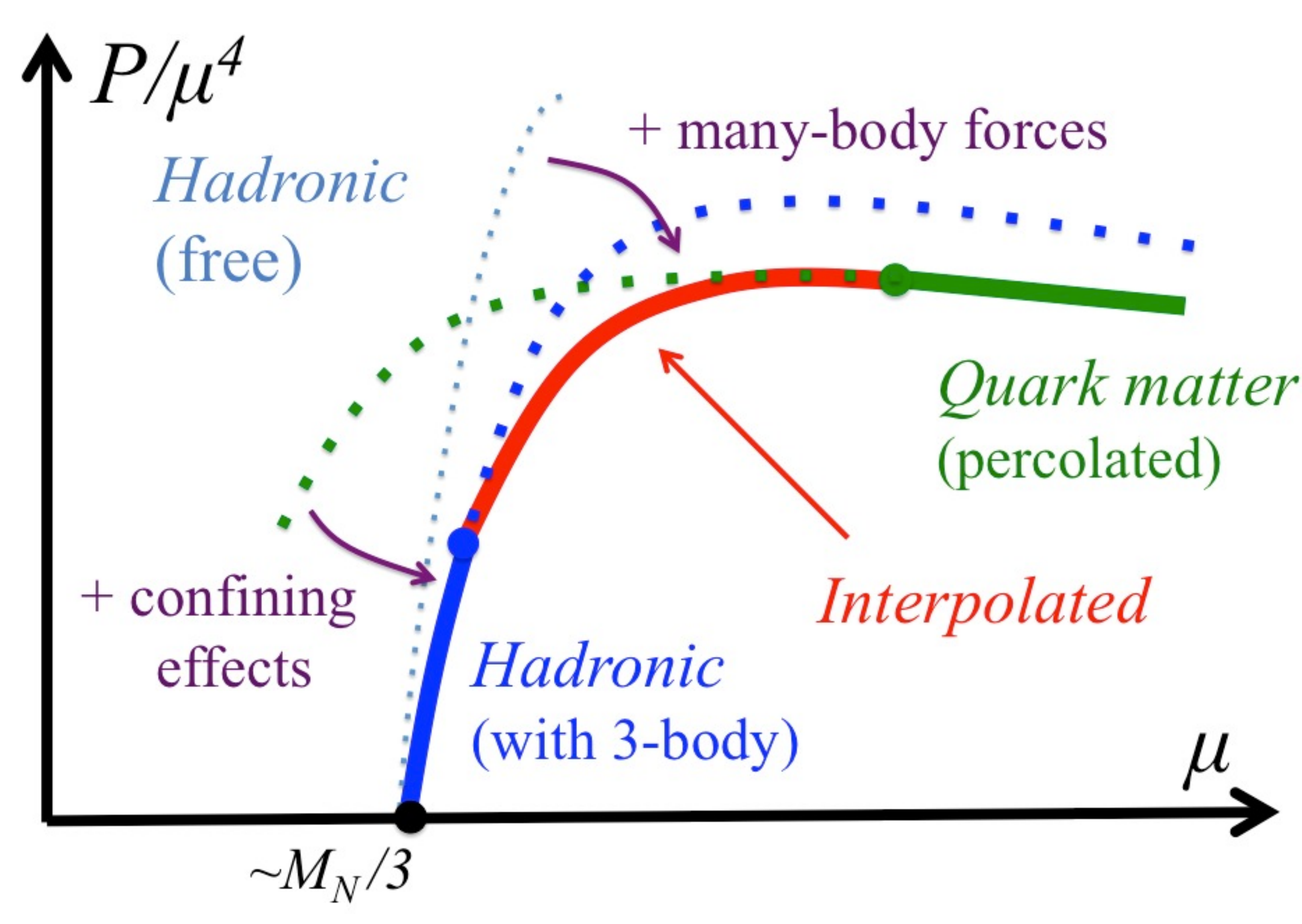}
}
\caption{Our expectation for the equation of state in finite density QCD. By interpolating the hadronic and percolated quark matter pressures, we expect that the 3-window pressure provides reasonably good description.}
\label{fig:inter_finitemu}  
\end{figure}

Now notice that if we trusted the extrapolated HRG or pQCD pressure as real, we would construct the hybrid equations of state in which we have pQCD gas at low temperature while HRG at high temperature! This senseless result tells us that one must be very careful when comparing the extrapolated equations of state. In contrast, if we stop using the HRG and pQCD gas pictures beyond the domains of their applicability, the domain $T_c < T < 2-3 T_c$ is left over. We can infer the pressure in such domain by interpolating the HRG and pQCD gas pressure, and then get the reasonable result; the 3-window construction for finite $T$ QCD equations of state works well. The domain $T_c < T < 2-3 T_c$ may be described as either strongly interacting HRG, or "semi-QGP" \cite{Pisarski:2000eq}, in which the quark-gluon plasma is subject to large non-perturbative effects. The finite density version of the semi-QGP is what we try to describe in the interpolated equations of state (Fig.\ref{fig:inter_finitemu}).
%

\subsection{The self-bound quark matter equations of state}
\label{selfbound_eos}

The thermodynamic ground state of a self-bound matter has nonzero number density $\partial P/\partial \mu |_{ \mu_{ {\rm th} } } \neq 0$ at threshold chemical potential $\mu_{ {\rm th}}$ such that $P( \mu_{ {\rm th} } )=0$. The self-bound quark matter hypothesis states that the absolute ground state of QCD matter is (percolated) strange quark matter \cite{Witten:1984rs}. If so, at low chemical potential the strange quark matter pressure is larger than the nucleonic one (see $P_{Q1}$ in Fig.\ref{fig:p_mu_hybrid}) and nuclear matter is just a meta-stable state. The decay from nuclear to strange quark matter is suppressed because of vastly different forms of these matters; wavefunctions do not quite overlap and it is hard to tunnel from one to another macroscopic state.
%

Strange quark matter equations of state can be made very stiff, if $\mu_{{\rm th}}$ and $n =\partial P/\partial \mu|_{ \mu_{ {\rm th} }}$ are small. In fact there are several examples of this sort that can pass the $2M_\odot$ constraint.
%

However, there are important qualifications, if one tries to use strange quark matter to explain the $2M_\odot$ neutron stars. As we saw in the last section, to get stiff equations of state we wish to make pressure slowly growing function of $\mu$. This condition requires $n=\partial P/\partial \mu|_{ \mu_{ {\rm th} }}$ to be small. However, if $n$ is too small or quark matter is too dilute, the confining effects are not negligible. The confining effects turn dilute strange quark matter into hypernuclear matter by packing three quarks into a nucleon or a hyperon. 

To be more concrete, we consider the bag model,
\beq
P(\mu) = \frac{\Nc\Nf}{12\pi^2} \mu^4-B\,,
\eeq
where we ignore the strange quark mass and $\alpha_s$ corrections. For this simplest version of the bag model, the star mass is  \cite{Lattimer:2000nx}
\beq
M_{\rm max} = 2.033 \times \left(\frac{56\, {\rm MeV/ fm^{3}} }{B} \right)^{1/2} \, M_\odot \,.
\eeq
So if $B\ge 56 {\rm MeV \cdot fm^{-3}}$, the self-bound quark matter can pass the $2M_\odot$ constraint. In this case $\mu_{{\rm th}} = 274\,{\rm MeV}$. But a question arises for the corresponding baryon density. It is
\beq
n_B(\mu_{ {\rm th} } ) = 0.273\,{\rm fm}^{-3} \simeq 1.7\, n_s\,,
\eeq
where the system is presumably too dilute to be free from the confining effects, with the fact that at this density purely nucleonic calculations do not find the serious problems yet. There are also analyses with $\alpha_s^2$ corrections to the bag model, but the equation of state compatible with the $2M_\odot$ constraint has the similar onset density \cite{Lattimer:2000nx}. 
%

For these reasons we are inclined to think of self-bound quark matter as a candidate of small, light stars. In this case, we can take the onset baryon density at $\mu_{{\rm th}}$ to be high enough to apply the percolation picture.
%

\section{A model of hadronic matter: APR}
\label{sec:APR}

For the 3-window construction it is necessary to choose a hadronic equation of state. For very dilute region at $n_B<0.5n_s$, we use the Skyrme-Lyon (SLy) equation of state for the crust \cite{Douchin:2001sv}. The crust part shares only small fraction of the neutron star mass, so the details are not essential for our discussions.  On the other hand, at $n_B \gtrsim 2n_s$ there are several issues on the validity of existing hadronic equations of state, so we will not try to directly describe such region. Then the remaining task is to choose one of equations of state for $0.5n_s<n_B<2n_s$, from the list available in the literature.
%

For $0.5n_s<n_B<2n_s$, the purely nucleonic equations of state are viable candidates. There are several nucleonic equations of state in the literature. They are constructed to reproduce the experimental data for symmetric nuclear matter at $n_B\simeq n_s$, but they are extrapolated to $n_B \sim 2n_s$ in different ways. Also, the differences can be already seen in the extrapolation to asymmetric nuclear matter, and they affect the neutron star radii by $\sim 1-3$ km  \cite{Lattimer:2000nx}.

In this work we use the Av18+$\delta v$+UIX$^*$ version of APR equation of state \cite{Akmal:1998cf}, that is based on variational many-body calculations for the non-relativistic potential models. The effective 2- and 3-body potentials are fixed by two nucleon scattering data below pion production threshold, spectra of light nuclei, and properties of symmetric nuclear matter at saturation density. (Nowadays more QCD-based derivation of such potentials is available in a systematic power counting in ChEFT \cite{Weinberg:1978kz,Epelbaum:2008ga,Bogner:2009bt}. Then those potentials are combined with the advanced many-body calculations; for recent progress, see Refs.\cite{Gezerlis:2014zia,Gandolfi:2015jma,Gandolfi:2011xu}.) 

The APR equation of state contains the first order phase transition for any proton fraction. The phase transition is interpreted as the pion condensation through the two-body correlation functions. For the $\beta$-equilibrium matter, the transition occurs at $n_B\simeq 1.4n_s$. (For symmetric nuclear matter, the transition density is higher, $n_B\simeq 2n_s$.) Although the strength of the phase transition is not too significant, we will see that this small amount of softening already introduces difficulties on the construction of stiff equations of state at high density.

Typically, the non-relativistic potential models provide softer equations of state at $n_B= 1-2n_s$ than mean field type calculations. The mean field type hadronic equations of state are rather stiff at low density and can be combined with stiff quark equations of state at high density (see the discussions around Fig.\ref{fig:p_mu_hybrid_eg2}). In fact there are several hybrid equations of state of this type which pass the $2M_\odot$ constraint \cite{Klahn:2006iw,Bonanno:2011ch,Benic:2014jia}. On the other hand, with the non-relativistic model pressure at low density, the construction of stiff equations of state becomes more difficult problem \cite{Alford:2004pf}, especially when we use the conventional hybrid construction.

Several nucleonic equations of state, which are constrained by the properties of symmetric matter at saturation density, can show considerable differences when they are extrapolated to asymmetric matter region. It is established that the neutron star radii are strongly correlated with the pressure at $n_B = 1-2n_s$ which is sensitive to the symmetry energy and its density dependence \cite{Lattimer:2000nx,Gandolfi:2011xu}. For mean field models the radii are typically $R=13-16$ km, while in the potential model $R=10-13$ km. The recent estimates of the neutron star radii from the observational data seem to favor rather small radii of $9-13$ km \cite{Steiner:2010fz,Steiner:2012xt,Ozel:2010fw,Ozel:2015fia,Guillot:2014lla,Heinke:2014xaa} with an exception suggesting $R>14\,{\rm km}$ \cite{Suleimanov:2010th}, although all these analyses contain some model assumptions. If these small radius estimates remain true, then we have to construct equations of state which are soft at $n_B=1-2n_s$ and becomes stiff at $n_B>2n_s$. The soft equations of state at $n_B=1-3n_s$ also seems to be favored by the heavy ion collision data by Danielewicz et al. \cite{Danielewicz:2002pu}.

While the nucleonic equations of state are safely applicable only up to $n_B\sim 2n_s$, the behaviors at low density already put severe constraints on high density equations of state through the boundary conditions, thermodynamic and causality constraints. Using the APR at low density, we need to combine the soft low density hadronic equations of state with stiff dense equations of state. Such combinations make the $2M_\odot$ constraints even severer. In the following sections we try to construct such equations of state.
%

\section{Modeling percolated quark matter}
\label{sec:model}

In this section we will introduce a schematic model for quark matter descriptions, and examine the impact of various effective interactions on the equations of state.
%

\subsection{Quark model}
\label{Quark_model}

At $n_B<10^2 n_s$, pQCD is not safely applicable and we must use some effective model for non-perturbative QCD. Our guideline toward proper effective models is hadron spectroscopy and nuclear physics. We consider the following schematic Hamiltonian\footnote{The Hamiltonian in this paper is a bit simpler than that of Ref.\cite{Kojo:2014rca} in which the anomaly coupling between chiral and diquark condensates is also included. The effect of such coupling can be mostly absorbed by changing the value of ($G_V,H$), so we will stick to the simpler version of the model.}:
\beq
\calH \sim \calH_{ {\rm NJL} } + \calH_{ {\rm conf} }^{ {\rm 3q\rightarrow B} } + \calH^{ {\rm mag.} }_{ {\rm OGE} } + \calH_{ {\rm BB} } + \cdots \,.
\eeq
Let us clarify physics which we wish to express. 

$\calH_{ {\rm NJL} }$ is the standard NJL model which is responsible for the chiral symmetry breaking/restoration and structural change of the Dirac sea and quark bases. It is given by
\beq
&& \calH_{ {\rm NJL} } 
= \overline{q} (\rmi \gamma_0 \vec{\gamma}\cdot \vec{\partial} + m -\mu \gamma_0)q 
\nonumber \\
&& - \frac{G_s}{2} \sum^8_{i=0} \left[ (\overline{q} \tau_i q)^2 + (\overline{q} \rmi \gamma_5 \tau_i q)^2 \right] 
+ 8 K ( \det\,\!\!_{\rm f} \bar{q}_R q_L + \mbox{h.c.}) \,,\nonumber \\
\eeq
where $m={\rm diag.}(m_u,m_d,m_s)$ is the current quark mass matrix, and $\tau_0=(2/\Nf)^{1/2}$ and $\tau_i (i=1,\cdots,8)$ are the Gell-Mann matrices for flavors. The last term is Kobayashi-Maskawa-'tHooft interaction for the $U_A(1)$ anomaly \cite{Kobayashi:1970ji}.
%

$\calH_{ {\rm conf} }^{ {\rm 3q\rightarrow B} }$ expresses the confining effects that trap 3-quarks into a baryon. This term will be essential if we try to describe hadronic matter microscopically. But right now we are not aware of the proper description in the practical models. Thus we will restrict the use of the model to the percolated domain, and then ignore $\la \calH_{ {\rm conf} }^{ {\rm 3q\rightarrow B} } \ra$ assuming its contribution to be small.
%

$\calH_{ {\rm OGE} }$ is a model for the one-gluon exchange contribution which is relatively short-range. We focus on the color-magnetic component:
\beq
\calH^{ {\rm mag.} }_{ {\rm OGE} } 
&=& - H \!\sum_{A,A^\prime = 2,5,7} \!
\big[ \left(\overline{q} \rmi \gamma_5 \tau_A \lambda_{A^\prime} C \overline{q}^T \right) \left(q^T C \rmi \gamma_5 \tau_A \lambda_{A^\prime} q \right)   \nonumber   \\
	       && \hspace{13mm} + \left(\overline{q} \tau_A \lambda_{A^\prime} C \overline{q}^T \right) \left(q^T C \tau_A \lambda_{A^\prime} q \right) \big] \,,
\eeq
where $\lambda_A$ are the Gell-Mann matrices for colors. Here we kept only the attractive channel: color-flavor-spin antisymmetric part. In hadron spectroscopy, it is responsible for the $N-\Delta$ splitting in the context of the constituent quark models. This interaction is called diquark-diquark interactions in researches of the color-superconductivity. 
%

$\calH_{ {\rm BB} }$ is inspired from baryon-baryon interactions. This is responsible for the quark version of the Walecka model \cite{Walecka:1974qa} type descriptions. Since we have already included the $\sigma$-meson part in the NJL model, we add only the $\omega$ and $\phi$ meson-like contributions or hard core type repulsive contributions \cite{Kunihiro:1991qu}:
\beq
\calH_{ {\rm BB} } = \frac{G_V}{2} (\overline{q} \gamma^\mu q)^2 \,.
\eeq
In principle we should include more general set of effective interactions for other channels. We will keep only the vector part for simplicity.
%

With this schematic Hamiltonian at hand, we will impose the constraints from charge and color neutrality, and $\beta$-equilibrium conditions  \cite{Iida:2000ha}. For these conditions, we introduce three chemical potentials (Lagrange multipliers) and add the following terms to the Hamiltonian:
\beq
\Delta \calH_{ {\rm const} }
= - \mu_Q (q^\dagger Q q - l_i^\dagger l_i ) - \mu_3 q^\dagger \lambda_3 q - \mu_8 q^\dagger \lambda_8 q \,,
\eeq
where $(\mu_Q, \mu_3, \mu_8)$ are chemical potentials for electric charge, the third and eighth components of the color matrices, respectively. (The other color charge densities automatically vanish for the condensates to be considered \cite{Buballa:2005bv,Abuki:2005ms}.) $Q$ is the electric charge matrix $Q={\rm diag.}(2/3, -1/3,-1/3)$, and $l_i (i=e,\mu)$ are electron and muon fields.
%

The Hamiltonian will be treated within the mean field approximation. The mean fields for the chiral condensates and quark number density are
\beq
\sigma_i & = & \la\overline{q}_i q_i \ra   ,  \hspace{3mm}  \hspace{3mm}  n =\sum_{i=1}^3 \la q^\dagger_i q_i \ra \,.
\eeq
Below we write $(\sigma_1, \sigma_2, \sigma_3)=(\sigma_u, \sigma_d, \sigma_s)$ for later convenience.  For the diquark mean fields, we write
\beq
d_i &= \la q^T C \rmi \gamma_5 R_i q \ra \,,
\eeq
where 
\beq
\left( R_1, R_2, R_3 \right) \equiv \left( \tau_7 \lambda_7, \tau_5 \lambda_5, \tau_2 \lambda_2 \right) \,.
\eeq
With these definitions, the diquark condensates $(d_1, d_2, d_3)$ correspond to $(ds, su, ud)$ quark pairings, respectively.
%

The thermodynamic potential may be computed from the mean field particle propagators in terms of these mean fields; the inverse of the propagator $S(k)$ can be read from the mean field Hamiltonian in the Nambu-Gor'kov bases \cite{Abuki:2005ms,Ruester:2005jc}, 
\beq
S^{-1} (k) = \begin{pmatrix}
		~ \slashed{k} - \hat{M} + \hat{\mu} \gamma^0 ~&~ \rmi \gamma_5 \Delta_i R_i ~\\
		~ - \rmi \gamma_5 \Delta^\ast_i R_i ~&~ \slashed{k} - \hat{M} - \hat{\mu} \gamma^0 ~
	     \end{pmatrix}   ,   \label{eq:Sinv}
\eeq
where the effective mass matrix has diagonal elements
\beq
M_i = m_i - 2 G_s \sigma_i + K |\epsilon_{ijk}| \sigma_j \sigma_k \,,
\eeq
while the three diquark pairing amplitudes,
\beq
\Delta_i = -2 d_i H    ,
\eeq
and the effective chemical potential matrix,
\beq
\hat{\mu} = \mu - G_V n +  \mu_3 \lambda_3 +\mu_8 \lambda_8 + \mu_Q Q \,,   \label{eq:mu_eff}
\eeq
are color- and flavor-dependent.
%

The thermodynamic potential for the quark part now reads
\beq
\Omega_q^{ {\rm bare} } & = & - 2 \sum^{18}_{j=1} \int^\Lambda \!\!\frac{d^3 \bd{k}}{(2\pi)^3} 
\left[ \frac{ |\epsilon_j| }{2} +T \ln \left( 1 + e^{- |\epsilon_j | /T }  \right) \right] 
\nonumber \\
&& \!\!
+ \sum^3_{i=1} 
\left[ G_s \sigma^2_i + H  |d_i|^2 \right] - 4 K \sigma_1 \sigma_2 \sigma_3 
- \frac{G_V}{2} n^2 ,~~~~~
\eeq
where the single particle contributions contain $18$ independent eigenvalues,
and $\Lambda$ is the UV cutoff for the Dirac sea contribution.
%

The "zero" of the renormalized thermodynamic potential is set by 
\beq
\Omega_q (\mu,T) \equiv \Omega_q^{ {\rm bare} } (\mu,T) - \Omega_q^{ {\rm bare} } (\mu=T=0)\,. 
\eeq
This choice of zero is demanded by the fact that the cosmological constant be extremely small.
%

The lepton contributions are standard. We add the renormalized (after subtraction of the vacuum energy)
thermodynamic potential ($l=e,\mu$)
\beq
\hspace{-0.3cm}
\Omega_l = -2 T\sum_{\lambda=\pm}  \int \frac{d^3 \bd{k}}{(2\pi)^3} \,\ln \left(1 + e^{-(E_{l} + \lambda \mu_Q)/T} \right) \,,
\eeq
with $E_{l} = \sqrt{\bd{k}^2 + m^2_l}$, and where we recall that the electron chemical potential is $\mu_{l^-} = - \mu_Q$.
%

Writing the total thermodynamic potential as $\Omega = \Omega_q + \Omega_e +\Omega_\mu$, 
the thermodynamic state of the system is determined minimizing the free energy 
with respect to the seven condensates \{$\sigma_i$,$d_i$,$n$\} under the neutrality conditions
\beq
n_{Q, 3, 8} = - \frac{\, \partial \Omega \,}{\, \partial \mu_{Q, 3, 8} \,}  = 0 ,  \label{eq:constraint_1}
\eeq
which yields the ``gap equations",
\beq
0 = - \frac{\, \partial \Omega \,}{\, \partial \sigma_i \,} = -\frac{\, \partial \Omega \,}{\, \partial d_i \,} \,,
~~~~ n = -  \frac{\, \partial \Omega \,}{\, \partial \mu \,} \,.
\eeq
Below we solve these self-consistent equations using the method outlined in~\cite{Abuki:2005ms,Ruester:2005jc}. 
%

%
\subsection{The impact of each interaction}
\label{impact_int}

Our schematic model contains a number of parameters. For the NJL part, we fix the parameters $(\Lambda, m, G_s, K)$ by the vacuum phenomenology\footnote{Strictly speaking, these parameters in medium can be smaller than the vacuum values. But as we will see, the neutron star constraints favor the values as large as the vacuum one. We will come back to this point in Sec.\ref{sec:gluons}. }. To be specific, we use the set provided by Hatsuda and Kunihiro \cite{Hatsuda:1994pi}:
\beq
&&\Lambda = 631.4\, {\rm MeV}\,,
~~~~~ G_s\Lambda^2 = 3.67\,,
~~~~~ K\Lambda^5 = 9.29 \,,
\nonumber \\
&& m_{u,d} = 5.5\,{\rm MeV}\,,
~~~~~ m_s =135.7\,{\rm MeV}\,.
\eeq
This set gives $M_{u,d}=336\,{\rm MeV}$ and $M_s=528\,{\rm MeV}$. For other choices of parameters, see \cite{Buballa:2003qv,Rehberg:1995kh,Lutz:1992dv}.
%

The other parameters, $H$ and $G_V$, are not constrained by the vacuum phenomenology. But from the viewpoint of the renormalization group, ($H,G_V$) are naturally the order of $G_s$. The standard choice is $(H,G_V)$ = $(0.75,0.50)G_s$ which is based on the Fierz transformation of the colored current-current interaction, $\sim (\bar{q}\gamma_\mu \lambda_a q)^2$. But this is just a tree level relation. The factor can easily be changed by the quantum corrections. Therefore we will consider the wider range for these parameters.
%

\subsubsection{The color magnetic interaction: $H$}
\label{diquark}

\begin{figure}
\begin{minipage}{1.0\hsize}
\hspace{0.5cm}
\includegraphics[width = 0.85\textwidth]{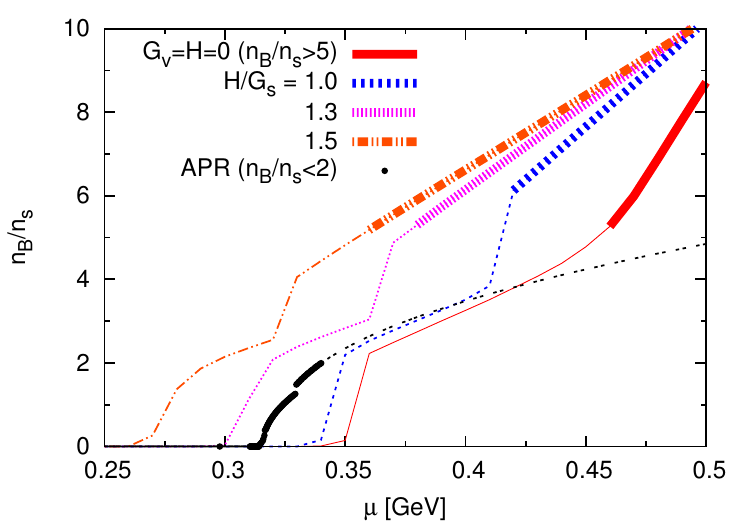}
\end{minipage}\\
\begin{minipage}{1.0\hsize}
\hspace{0.5cm}
\includegraphics[width = 0.85\textwidth]{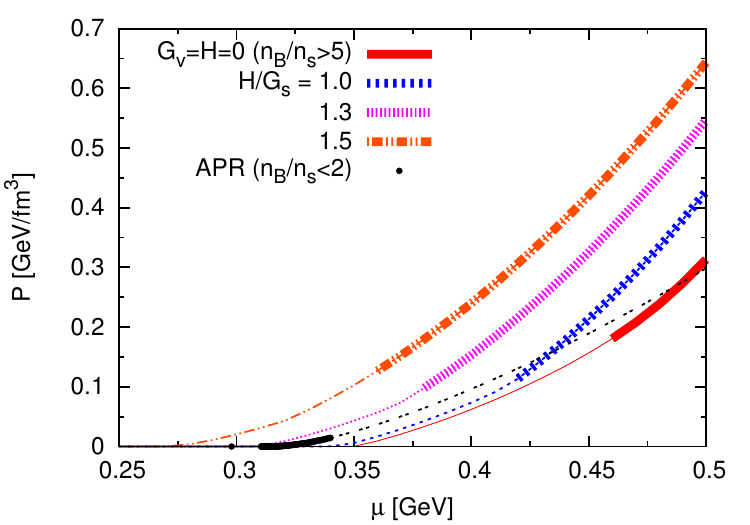}
\end{minipage}
\caption{
\footnotesize{Normalized baryon density $n_B/n_s$ (top panel) and pressure $P$ (bottom panel) as functions of quark chemical potential $\mu$ for several NJL parameter sets ($G_V=0$ and several values for $H$) and APR. The only trustable region is shown with the bold lines ($n_B<2n_s$ for APR and $n_B>5n_s$ for NJL).  At $H>G_s$, the baryon density and pressure in the NJL model begin to appear at lower chemical potential than $M_N/3 \simeq 313\,{\rm MeV}$ where non-confining models are not trustable.}
}\label{fig:gv00_H_vary_mu-n_p}
\vspace{-0.2cm}
\end{figure}

\begin{figure}
\hspace{0cm}
\resizebox{0.45\textwidth}{!}{%
  \includegraphics{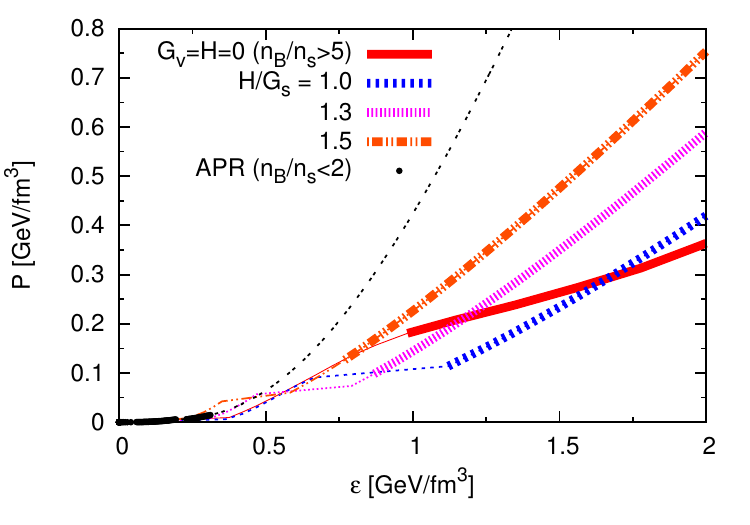}
}
\caption{The pressure as a function of energy density. While $H>G_s$ introduces the first order phase transition and the associated softening, at larger density the pairing effects eventually make the equations of state stiffer than those without the pairing. (See also discussions around Fig.\ref{fig:p_mu_e_eg1}.)}
\label{fig:gv00_H_vary_e-p}  
\end{figure}

The standard choice of $H$ is $0.75\, G_s$, while the choice $H=G_s$ is called strong coupling. On the other hand, the choice $H>G_s$ has been avoided, presumably because such large $H$ yields the diquark condensates at density even lower than nuclear saturation density (Fig.\ref{fig:gv00_H_vary_mu-n_p}). In this case the 2SC phase ($\Delta_{ud}\neq 0, \Delta_{ds}=\Delta_{su}=0$) appears first and then CFL phase ($\Delta_{ud}, \Delta_{ds},\Delta_{su}\neq 0$), but no normal quark matter phase.
%

But the appearance of diquark condensates in very dilute regime might be simply an artifact of non-confining models, for which the overlooked corrections are significant at low density. Therefore, one should not take the problem literally and cannot use such problem to constrain the value of $H$ which will be also used at high density. If we could add the confining effects to the NJL model, the colored diquarks would be combined  with colored quarks to form baryons. In other words, in dilute regime diquarks would exist only inside of baryons, without showing themselves in nuclear phenomenology.
%

Moreover, there is a physics motivation to take $H>G_s$. Let us first note that in the NJL model the constituent quark mass for light quarks is $\simeq 336\,{\rm MeV}$, larger than $M_N/3  \simeq 313\, {\rm MeV}$. Thus the NJL pressure tends to appear at higher chemical potential than the hadronic pressure, and as the consequence these pressures look very different. But microscopically we should be able to describe hadronic matter in terms of quark degrees of freedom, so there must be some effects overlooked. One of the effects to be implemented is attractive forces which reduce the average of quark energy to the smaller energy than $M_N/3$ as seen in the $N-\Delta$ splitting where the nucleon earns $\simeq 150\,{\rm MeV}$ energy reduction. We will use the color-magnetic interaction to describe this sort of effect, and our purpose is achieved by taking a large value, $H >G_s$.
%

For a choice $H>G_s$, the quark matter pressure of {\it non-confining} models start to appear at $\mu<M_N/3$ and can be larger than the hadronic pressure. The artificial excess of pressure should be suppressed by inclusion of the missing confining effects; packing quarks in the finite domain and the color flux tubes among quarks both cost energy. In this way of thinking, it is rather natural to have excess of low density pressure in the non-confining model. This picture motivates us to use $H>G_s$. The value of $H$ is correlated with other parameters. If, for instance, the constituent quark mass appeared closer to (or smaller than) $M_N/3$, we could take smaller $H$; if the quark mass appeared larger, we had to take larger $H$ to achieve larger energy reduction. 
%

Finally we discuss the softening/stiffening induced by the color magnetic interactions. We have been accustomed to the examples in which the condensation phenomena soften the equation of state. This idea is not always true, however, as seen in Fig.\ref{fig:gv00_H_vary_e-p}. Certainly it is always true that the first order phase transitions soften equations of state, and most of model studies have emphasized only this aspect. However, it is quite possible that the stiffness in the condensed phases grows fast and eventually exceeds that in the normal phase. This can be seen in Fig.\ref{fig:gv00_H_vary_mu-n_p}, where the diquark correlation induces overall shift of pressure curves toward lower chemical potential region, at least in the large $\mu$ region. We have already discussed that such shift stiffens the equation of state (see discussions around Fig.\ref{fig:p_mu_e_eg1}).
%

Therefore, the question relevant to neutron star phenomenology is at which density condensed phases become stiffer than the normal phase. For small values of $H$ such reversal does not occur at density of interest, and this statement is consistent with the previous works. But for sufficiently large $H$, such reversal can occur at low density as seen in Fig.\ref{fig:gv00_H_vary_e-p}. At large $H (>1.5)$, the system directly goes to the condensed phase without experiencing the normal phase, and the condensed phase is stiffer than the normal phase from low to high densities. These discussions show that whether condensed phases stiffen the equations of state or not is a quantitative issue.

\subsubsection{The repulsive density-density interaction: $G_V$}
\label{sec:vectorint}

\begin{figure}
\hspace{0cm}
\resizebox{0.45\textwidth}{!}{%
  \includegraphics{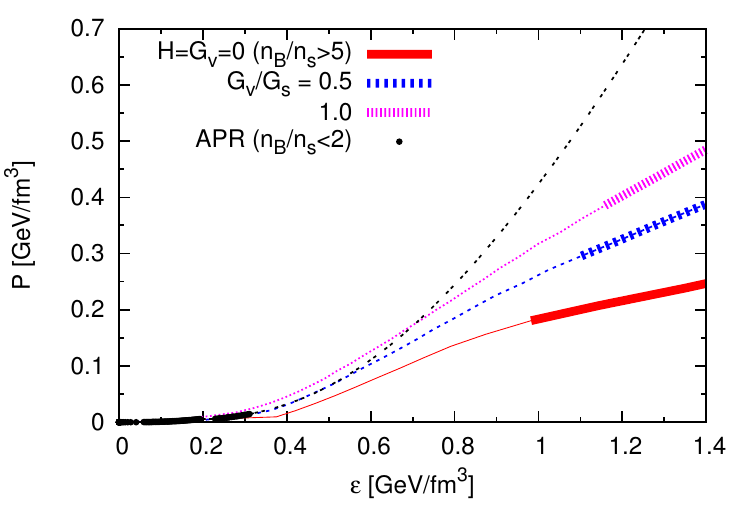}
}
\caption{The pressure as a function of energy density. Large $G_V$ leads to stiffer equations of state.}
\label{fig:gv_vary_H000_e-p}  
\end{figure}

\begin{figure}
\begin{minipage}{1.0\hsize}
\hspace{0.5cm}
\includegraphics[width = 0.85\textwidth]{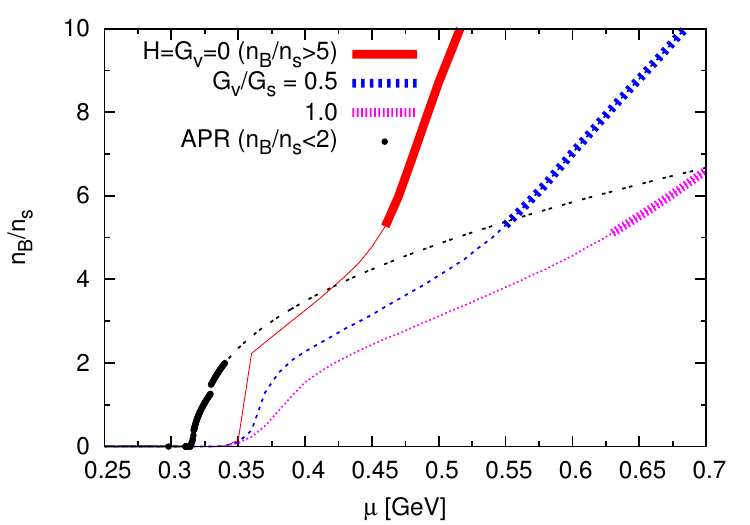}
\end{minipage}\\
\begin{minipage}{1.0\hsize}
\hspace{0.5cm}
\includegraphics[width = 0.85\textwidth]{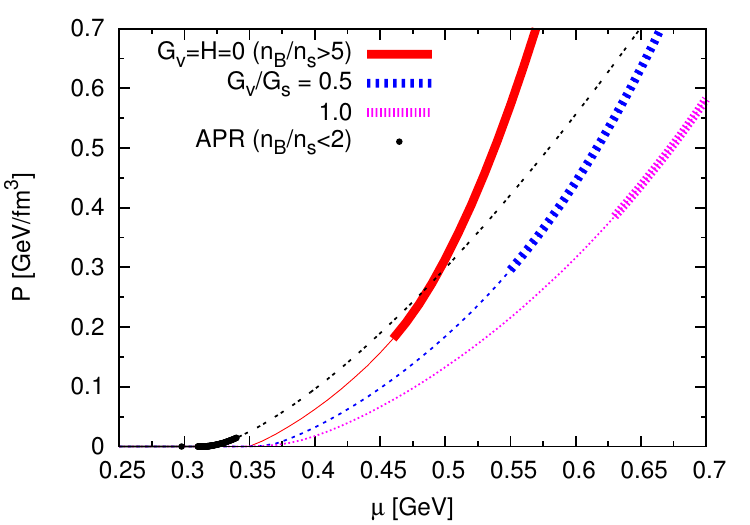}
\end{minipage}
\caption{
\footnotesize{Normalized baryon density $n_B/n_s$ (top panel) and pressure $P$ (bottom panel) as functions of quark chemical potential $\mu$ for several NJL parameter sets ($H=0$ and several values for $G_V$) and APR.}
}\label{fig:gv-vary_H000_mu-n_p}
\vspace{-0.2cm}
\end{figure}

The repulsive density-density interaction plays very important roles to stiffen equations of state (Fig.\ref{fig:gv_vary_H000_e-p}). It is intuitively clear that the repulsive force tries to prevent number density from increasing (Fig.\ref{fig:gv-vary_H000_mu-n_p}); the repulsion increases the chemical potential which is necessary to increase the density and also pressure. The resulting $P(\mu)$ curves are slowly growing functions that are characteristic of stiff equations of state.
%

Another important consequence of the vector interaction is that it smoothes out phase transitions \cite{Kitazawa:2002bc}. Because baryon density changes slowly as chemical potential increases, the transition driven by baryon density also occurs slowly. As for the chiral transition at $G_V=0$, the chiral restoration is triggered by the increase of baryon density, while the resulting reduction of the effective mass further accelerates the increase of baryon density. These two effects drive each other and the changes occur very rapidly, resulting in the first order phase transition. On the other hand, increasing $G_V$ brakes the increase of baryon density, so that the chiral restoration occurs only slowly. At large enough $G_V$, the vector interaction turns the first order phase transition into crossover; in general, introducing the repulsive density interactions shifts the chiral restoration line toward higher chemical potential region, and, as a result of the weaker strength of the phase transition, the location of the critical end point moves toward low temperature and eventually disappears at large $G_V$. In the standard NJL model, this occurs at $G_V\simeq 0.4G_s$.
%

This feature suggests us an interesting interpretation on the lattice data for chiral restoration. The lattice Monte-Carlo calculations indicate that the curvature of the chiral restoration line at $\mu=0$ is unexpectedly flat \cite{Kaczmarek:2011zz}, and the chiral restoration line seems to be separated from the line characterizing the onset of baryon density. One of the simplest explanations on this behavior is to assume the existence of repulsive density-density interactions. In NJL-like models, this effect can be expressed by the vector interaction. Seminal works \cite{Bratovic:2012qs,Contrera:2012wj} used this feature to constrain the value of $G_V$ in the Polyakov-NJL (PNJL) model \cite{Fukushima:2003fw}, and they found that $G_V/G_s=0.5-1.5$ is reasonable range not to deviate from the curvature in the lattice data by $\sim 50\%$. (On the other hand, the authors in Refs.\cite{Ferroni:2010xf,Steinheimer:2010sp} compared the quark number susceptibility in the effective models with the lattice result at $T>T_c$, and argued that $G_V$ should be small enough not to suppress the susceptibility too much. In Sec.\ref{sec:gluons}, we will discuss this case further in the context of gluon dynamics.) We will also see that a large $G_V$ is necessary to get enough stiffness to sustain the $2M_\odot$ neutron stars.
%

\section{The 3-window equations of state}
\label{sec:3-window}

In this section we construct 3-window equations of state by interpolating the APR and NJL equations of state. Then we calculate a neutron star mass to select out the equations of state which pass the $2M_\odot$ constraint. The $M$-$R$ relation for non-rotating, spherically symmetric stars can be calculated through TOV equation,
\beq
\frac{\, \rmd P \,}{\rmd r} = - \frac{G_N (\varepsilon + P) ({\cal M} + 4 \pi r^3 P)}{ r(r-2 G_N {\cal M} ) }
\eeq
where $G_N$ is the Newton constant, and
\beq
{\cal M} (r) = \int_0^r \rmd r' 4\pi r'^2 \varepsilon(r') \,.
\eeq
With the central baryon density $n_B^c$ (or central energy density $\varepsilon_c$) as the boundary condition at $r=0$, we use the QCD equations of state and the TOV equations.
The star radius $R$ is determined as
\beq
P(r=R; n_B^c) = 0\,,
\eeq
and the star mass is given by
\beq
{\cal M} (n_B^c) = {\cal M} (R; n_B^c) \,.
\eeq
For each equation of state we have a $M$-$R$ curve as a function of $n_B^c$.
%


\subsection{The interpolated equations of state}
\label{sec:interpolated}
\begin{figure}
\hspace{0cm}
\resizebox{0.45\textwidth}{!}{
  \includegraphics{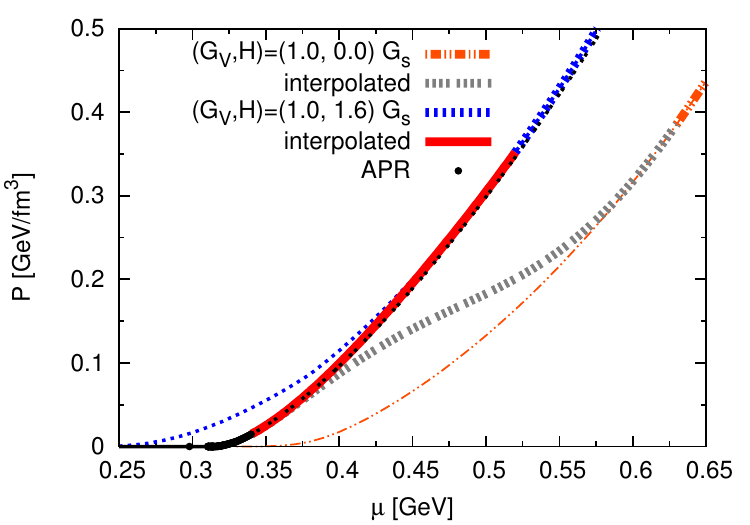}
  }
\caption{The 3-window pressures as functions of $\mu$ for the parameter set $(G_V,H)=(1.0, 0.0)G_s$ (set I) and $(1.0, 1.6)G_s$ (set II). The set I gives unphysical interpolation: the curve contains the inflection point. At $n_B>2n_s$, the pressure with the set II almost coincides with the APR curve. The bold lines are used for APR at $n_B<2n_s$ and NJL at $n_B>5n_s$, and their extrapolations are shown in thiner lines but with the same colors.}
\label{fig:eg_mu-P}  
\end{figure}
%
\begin{figure*}
\begin{minipage}{0.9\hsize}
\vspace{-0.3cm}
\hspace{-0.3cm}
\includegraphics[width = 1.0\textwidth]{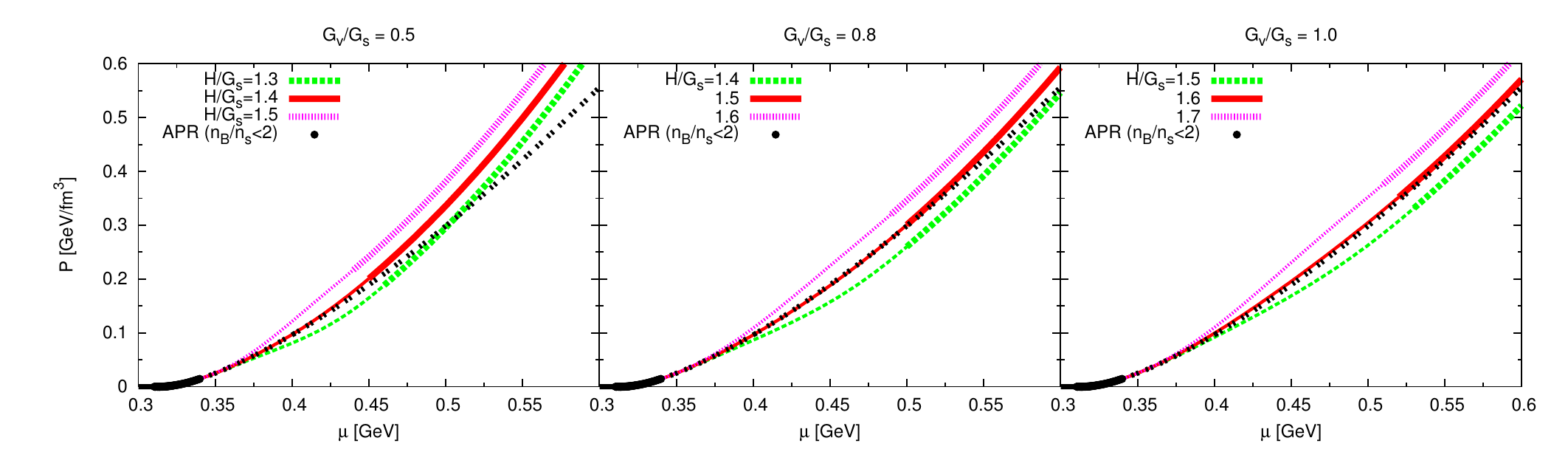}
\end{minipage}
\vspace{-0.2cm}
\caption{$P$ v.s. $\mu$ for several vector couplings, $G_V/G_s=0.5,0.8,1.0$. For each vector coupling, we take the value of $H$ around which we can find physical interpolated equations of state. The curves with inflection points must be excluded. The bold lines are used for APR at $n_B<2n_s$ and for NJL at $n_B>5n_s$.
\footnotesize{
}
\vspace{-0.3cm}
}\label{fig:gv05_08_10_mu-p}
\end{figure*}
\begin{figure*}
\begin{minipage}{0.9\hsize}
\hspace{-0.3cm}
\includegraphics[width = 1.0\textwidth]{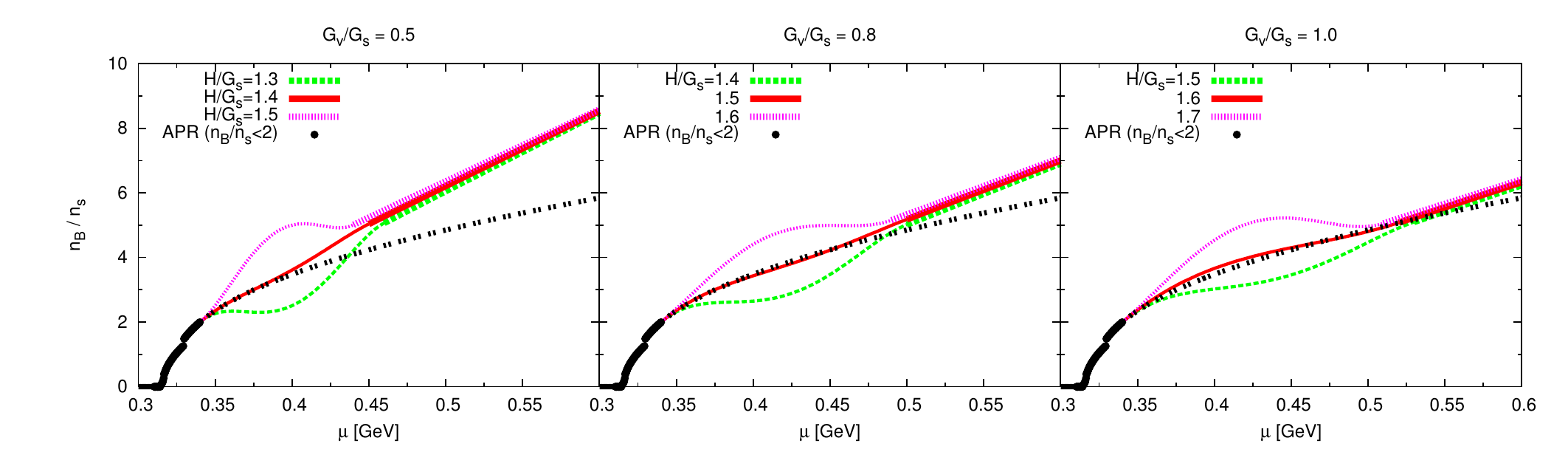}
\end{minipage}
\vspace{-0.2cm}
\caption{
\footnotesize{$n_B/n_s$ v.s. $\mu$. The curves with decreasing behaviors must be rejected. At high density, the curves with different $H$ almost coincide, indicating that increasing $H$ induces overall shift of pressure curves without changing the slopes much.
}
\vspace{-0.3cm}
}\label{fig:gv05_08_10_mu-n}
\end{figure*}

In this work we use a polynomial function to interpolate the APR and NJL equations of state. For the matching conditions at low and high densities, we demand $P(\mu)$ to smoothly join the APR and NJL pressures up to the second derivatives; $P$, $n$, and the quark number susceptibilities are matched. Explicitly, first we define the chemical potentials at the boundaries as ($n_B=3n$)\footnote{To check the sensitivity of our results to the boundary densities, we examined results by varying $n_<$ as $1.5-3n_s$ and $n_>$ as $4-7n_0$. These changes do not introduce any substantial effects as far as the interpolation window is not taken to be too narrow. Also the effects of such changes can be easily absorbed by slightly changing the model parameters $(G_V,H)$.}
\beq
n_B(\mu_<) \equiv 2n_s\,,~~~~~~ n_B(\mu_>) \equiv 5n_s\,.
\eeq
The polynomial function is 
\beq
\calP(\mu) = \sum_{k=0}^5 c_k \mu^k \,,
\eeq
where we have six coefficients to satisfy three matching conditions at each boundary,
\beq
\frac{ \partial^j \calP \,}{ \partial \mu^j} \bigg|_{\mu_<} 
= \frac{ \partial^j P_{{\rm APR}} }{ \partial \mu^j} \bigg|_{\mu_<} ,
~~
\frac{ \partial^j \calP \,}{ \partial \mu^j} \bigg|_{\mu_>} 
= \frac{ \partial^j P_{{\rm NJL}} }{ \partial \mu^j} \bigg|_{\mu_>} ,
\eeq
where $j=0,1,2$.
%

As the form of the interpolating function suggests, here we are assuming the crossover behavior between hadronic and quark matter. But the 3-window construction itself does not necessarily demand such crossover; one can use the interpolating function which contains a kink which is characteristics of the first order phase transition, although its strength must be small enough to avoid significant softening. In such case the final form of the 3-window equations of state can be similar to those found in the conventional hybrid construction; the only difference in these two constructions are whether or not the extrapolated pressures are utilized explicitly.
%

The polynomial function apparently allows many different kinds of interpolated equations of state. Actually it is not so. Once we require the pressure curve to yield the physical speed of sound, the number of the candidates drastically decreases.
%

Let us first consider the 3-window equation of state with large $G_V$, in order to pass the $2M_\odot$ constraint. While larger $G_V$ certainly stiffens the equations of state, it also brings a trouble in sensible interpolations. As seen in Fig.\ref{fig:eg_mu-P}, the NJL pressure with large $G_V$ tends to appear at higher chemical potential region than the APR curve. As a consequence the NJL and APR curves cannot be connected without introducing the inflection points with which $c_s^2$ becomes negative. Thus we wish to increase $G_V$ to pass the $2M_\odot$ constraint but the aforementioned unphysical behavior prevents us from doing so.
%

The NJL pressure tends to appear at higher density because the NJL constituent quark mass is $M_q\simeq 336\,{\rm MeV}$, larger than the one-third of nucleon mass, $\simeq 313\,{\rm MeV}$, that gives the threshold chemical potential of APR. As we saw in the last section, as we turn on the attractive pairing forces among quarks, the NJL pressure shows overall shift toward the lower chemical potential region, approaching the APR pressure. For the $G_V=G_s$ case shown in Fig.\ref{fig:eg_mu-P}, we increased the value of diquark coupling until we found the sensible interpolated pressure. We found that for $G_V=G_s$, the strong diquark coupling of $H\simeq 1.6\,G_s$ is necessary to avoid the inflection point. The resulting 3-window pressure almost coincides with the APR pressure, although they gradually depart each other as density goes beyond $\sim 5n_s$. 
%

In Fig.\ref{fig:eg_mu-P} we can see the set $(G_V,H)=(1.0,1.6)G_s$ yields nonzero pressure at $\mu<M_N/3$. As we have emphasized in several places, at low density, the extrapolated pressure of {\it non-confining} models may be (or should be) larger than the APR, since the confining effects are not included yet. As we will see, the equations of state passing the $2M_\odot$ constraint and speed of sound commonly show this excess pressure at low density.
%

\begin{figure*}
\begin{minipage}{0.9\hsize}
\hspace{-0.3cm}
\includegraphics[width = 1.0\textwidth]{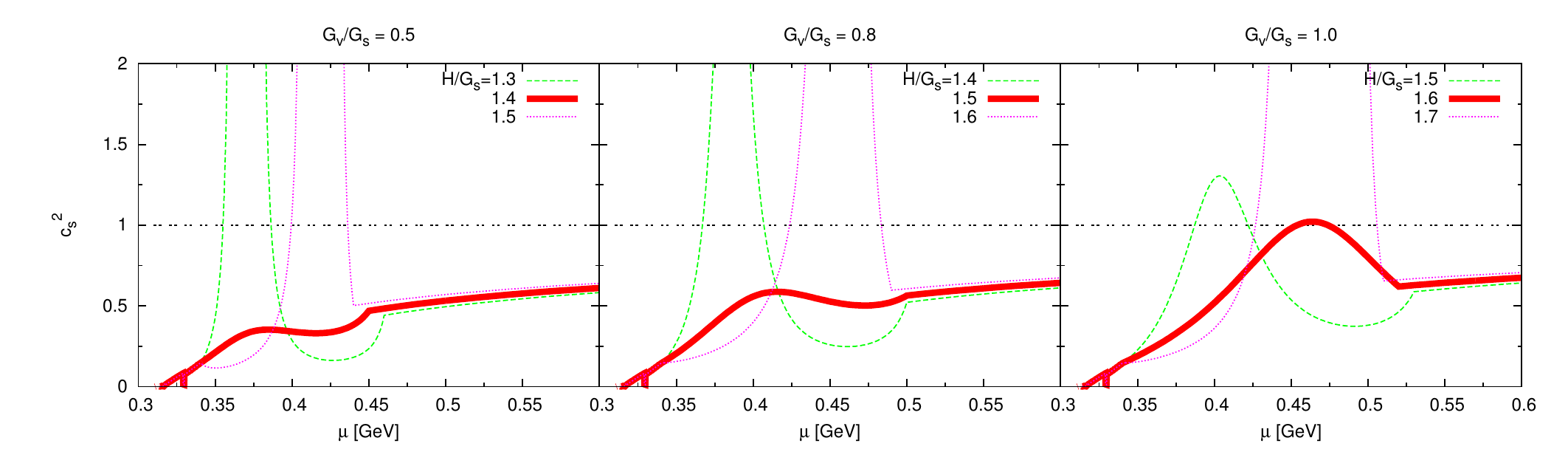}
\end{minipage}
\vspace{-0.2cm}
\caption{
\footnotesize{The (square of) speed of sound $c_s^2$ as a function of $\mu$. The negative $c_s^2$ exist but are not plotted. For each $G_V$, only one of curves (bold, red) has the physical speed of sound. (Strictly speaking, at $G_V=G_s$ the value of $c_s^2$ is slightly larger than $1$ and we need to take $H$ slightly smaller than $1.6$ to pass the causality constraint. This change does not affect our arguments.)
}
\vspace{-0.2cm}
}\label{fig:gv05_08_10_mu-cs2}
\end{figure*}
\begin{figure*}
\begin{minipage}{0.9\hsize}
\hspace{-0.25cm}
\includegraphics[width = 1.0\textwidth]{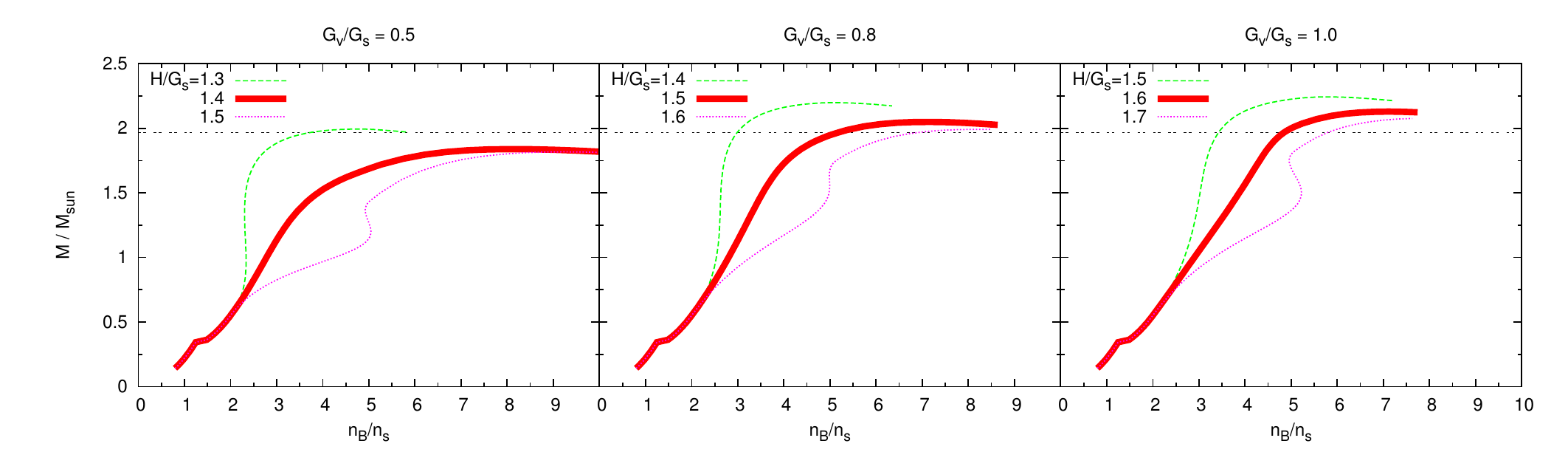}
\end{minipage}
\vspace{-0.2cm}
\caption{
\footnotesize{The neutron star mass $M/M_\odot$ v.s. $n_B/n_s$ for the parameter set discussed in Fig.\ref{fig:gv05_08_10_mu-p}. Only bold (red) lines are physically acceptable. To pass the $2M_\odot$ constraint, the vector coupling needs to be sufficiently large, $G_V/G_s>0.5$.
}
\vspace{-0.3cm}
}\label{fig:gv05_08_10_n-m}
\end{figure*}

Now we move to more systematic studies in the parameter space of $(G_V,H)$. In Figs.\ref{fig:gv05_08_10_mu-p},\ref{fig:gv05_08_10_mu-n},\ref{fig:gv05_08_10_mu-cs2}, we plot $P$, $n_B/n_s$ and $c_s^2$, respectively, as functions of $\mu$. In Fig.\ref{fig:gv05_08_10_n-m}, we plot the neutron star mass as a function of the central number density, $n_B^c$. These results are given for $G_V/G_s=0.5, 0.8$ and $1.0$. As for the values of $H$, we searched the central value of $H$ so as to give sensible interpolation at given $G_V/G_s$, and then study the slight variations from it. Several remarks are in order:
%

(i) For sensible interpolations, the appropriate values of $H$ becomes larger as we increase $G_V$. For $G_V/G_s=0.5$,$0.8$, and $1.0$, the appropriate choices of $H/G_s$ are $1.4$, $1.5$, and $1.6$, respectively (we have not tried to fine-tune the values of $H$ more than the accuracy of $0.1$). To check whether the curves are physical or not, we examine the behaviors of number density and speed of sound in Figs.\ref{fig:gv05_08_10_mu-n} and \ref{fig:gv05_08_10_mu-cs2}.
%

(ii) The resulting pressure curves are similar to the APR pressure, indicating that the boundary conditions up to the second derivatives are considerably tight. The matching domain becomes wider for larger $G_V$, beyond such domain the NJL pressure gets softer than APR. It seems that $G_V/G_s=0.5$ is a bit too small to pass the $2M_\odot$ constraint, while $G_V/G_s=0.8$ and $1.0$ allows us to pass the constraint.
%

(iii) The conditions on the speed of sound appear to be quite tight. Because the APR is relatively soft at $n_B= 1-2n_s$, the stiff equations of state at high density side necessarily requires the rapid growth in $P(\varepsilon)$. With careless interpolation one easily finds the violation of the causality. 
%

(iv) As seen from Fig.\ref{fig:gv05_08_10_n-m}, a neutron star with the mass $M/M_\odot \gtrsim 0.7$ has the central density of $n_B>2n_s$, and requires information beyond purely nuclear matter. The equations of state at $n_B> 4n_s$ carry vital information about whether they can pass the $2M_\odot$ constraint.
%

\subsection{The mass-radius relations}
\label{sec:M-R}
\begin{figure}
\hspace{0cm}
\resizebox{0.45\textwidth}{!}{
\includegraphics{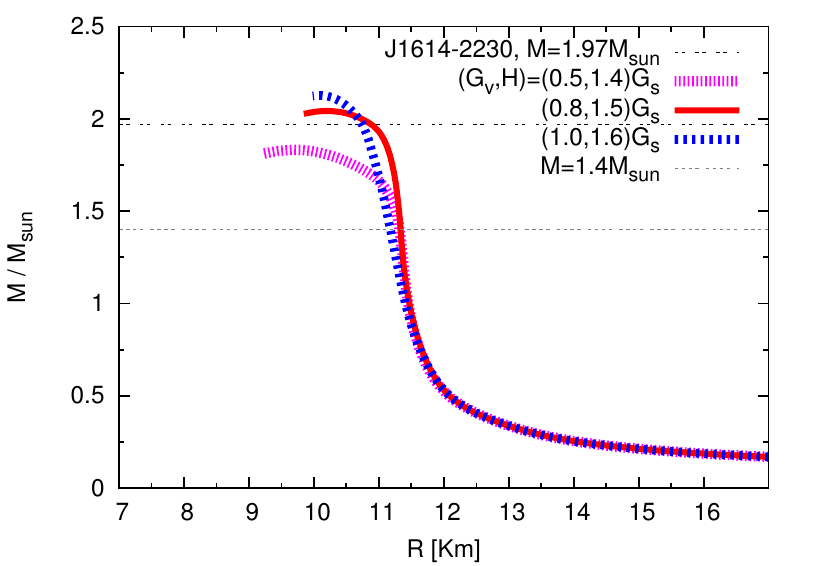}
}
\caption{The $M$-$R$ relation. From the equations of state displayed in Fig.\ref{fig:gv05_08_10_mu-p}, we picked out only physically acceptable ones. For canonical neutron stars with $M\simeq 1.4M_\odot$, the radii are about $R \simeq 11.2-11.5\,{\rm km}$, mainly determined by APR for $n_B\sim 2n_s$.}
\label{fig:gv05_08_10_M-R}  
\end{figure}

For the $M$-$R$ relations, we pick up only physical 3-window equations of state from Fig.\ref{fig:gv05_08_10_mu-p}: the parameter sets are $(G_V/G_s, H/G_s)=(0.5,1.4)$, $(0.8,1.5)$, and $(1.0,1.6)$. The results are shown in Fig.\ref{fig:gv05_08_10_M-R} together with the lines of $1.97 \pm0.04 M_\odot$ for J1614-2230 and of the canonical mass, $1.4M_\odot$. As we already discussed, larger $G_V$ leads to the larger neutron star mass. As for the neutron star radii, $R \simeq 11.2-11.5$ km at the canonical mass. 
%

The shape of a $M$-$R$ relation deduced from our 3-window result is similar to that from the APR, as expected from the behaviors of pressure. But we should notice that the degrees of freedom in the equations of state are quite different; while the APR contains only nucleonic degrees of freedom to very high density, the 3-window equations of state manifestly take into account the troublesome strangeness degrees of freedom. This difference should be crucial since the $2M_\odot$ neutron stars need information around $\sim 5n_s$ where strangeness is supposed to exist.
%

The consequences of our 3-window and conventional hybrid equations of state can be distinguished by the neutron star radii. In conventional hybrid construction, we tend to choose stiff hadronic equations of state at low density that enables stiff quark matter pressure to intersect with the hadronic one from below. With the stiff hadronic equations of state, the neutron star radii tends to be large, $R=13-16$ km. In contrast, in our 3-window construction we use soft hadronic equations of state at low density, and, at the same time, stiff quark matter equations of state at high density; the former is responsible for small radii of $\simeq 11$ km while the latter is used to yield the mass of $\simeq 2M_\odot$.
%

\section{Discussions}
\label{sec:discussions}

So far we have examined the interpolated pressure but have not discussed the physics in it. In this section we will infer the physics in the interpolated region. In the following we will discuss the hyperon puzzle and the nature of gluon dynamics at density relevant to neutron stars.
%

\subsection{On the hyperon problem}
\label{sec:hyperon}

In purely hadronic descriptions with hyperons, the many-body approaches for potential models such as Hartree-Fock \cite{Massot:2012pf}, Brueckner-Hartree-Fock \cite{Schulze:2011zza} give equations of state which encounter the hyperon softening around $n_B=2-3n_s$. The resulting maximal star mass is about $\sim 1.6 M_\odot$, smaller than the $2M_\odot$ constraint. The results indicate that for the $NN$ and $YN$ potentials ($N$ and $Y$ are nucleons and hyperons, respectively), not only repulsive two-body forces but also more-body repulsion are necessary to prevent matter from softening. Recent Quantum Monte Carlo calculations showed the big impact of the 3-body forces which have yet been determined empirically \cite{Lonardoni:2014bwa}.


The mechanisms proposed in the context of hadronic descriptions typically utilize the higher order effects such as many-body forces to delay the appearance of hyperons. But whenever the higher order effects have too much importance, there should arise the questions about the convergence. Then the relevant question to be answered is how they behave after summing up the higher order corrections.
%

To guess how the results change toward high density, let us consider the hyperon problem from the microscopic viewpoint. In quark model descriptions, the strangeness begins to appear when quark chemical potential reaches $M_s \simeq 0.5\, {\rm GeV}$ if we ignore the chiral restoration. In terms of baryon chemical potential, it amounts to the order of $\mu_B \sim 3M_s \sim 1.5\,{\rm GeV}$, considerably higher than the $\Lambda$ or $\Sigma$-masses, $m_{\Lambda, \Sigma} \sim 2 M_{u,d} +M_s \sim 1.1\,{\rm GeV}$. Thus the strangeness appears at relatively higher chemical potential  in the quark description than in the hadronic one. 
%

Let us consider why this is the case. Let us first use purely hadronic descriptions. For simplicity, we consider pure neutron matter with $\Lambda$ hyperons and ignore the interactions for the moment (therefore the numerics below should not be considered too seriously). In purely hadronic descriptions, $\Lambda$ at rest appears through a weak decay, $n(udd)\rightarrow \Lambda(uds)$, when the neutron Fermi energy $E_{nF} =\sqrt{m_n^2 +p_{nF}^2}$ reaches the $\Lambda$-mass. The corresponding neutron Fermi momentum is $\simeq 0.6\,{\rm GeV} \sim \lqcd$. Treating baryons as if elementary particles, the onset baryon chemical potential for $\Lambda$ is $\mu_B=m_\Lambda$.
%

The above description, however, ignores important aspects of the matter made of composite particles. At baryon chemical potential where the neutron Fermi momentum becomes the order of $\lqcd$, the phase space for $u$ and $d$-quarks is occupied to quark momenta of $\sim \lqcd$. In such situation, $\Lambda$ baryons, composite particles made of $u,d$ and $s$-quarks, cannot be free from the Pauli blocking effects acting on $u,d$ quarks. For instance, $\Lambda$ baryons {\it at rest} are dominantly described by quarks at low momenta, and their emergence is obstructed by already existing $u,d$ quarks in the neutron Fermi sea. While the $\Lambda$-baryons at sufficiently large momenta can escape from these Pauli blocking effects, such $\Lambda$ is too energetic and cannot be generated through the neutron weak decays. One can create $\Lambda$ at rest by replacing $d$-quark at low momenta with $s$-quark, but it increases the energy of the system by $\sim M_s-M_d$, thus cannot occur in the $\beta$-equilibrium. The only strange baryon that is perfectly free from the quark Pauli blocking is $\Omega^-(sss)$, but it appears only when $\mu_B$ reaches $m_\Omega \sim 1.6\,{\rm GeV} \sim 3M_s$. The last case roughly corresponds to the onset baryon chemical potential in quark matter descriptions. 
%

To be more concrete, consider a baryon in a constituent quark model, for example, of Isgur and Karl \cite{Isgur:1978xj}. The spatial part of the ground state baryon wavefunction at momentum $\vec{K}$ can be expressed by quark wavefunctions as ($\vec{P}=\vec{p}_1 + \vec{p}_2 + \vec{p}_3$)
\beq
\Psi^{{\rm spatial}}_{\vec{K}} (\vec{p}_1, \vec{p}_2,\vec{p}_3)
&=& \frac{(2\pi)^{3/2} }{\, \sqrt{V} \,} \, \delta(\vec{K}-\vec{P})
\nonumber \\
&&\times 
\left( \frac{ 4\sqrt{3}\pi}{\lambda^2} \right)^{3/2}
\, \rme^{- \frac{1}{2\lambda^2 } ( \vec{p}_\rho^2 + \vec{p}_\kappa^2 ) }\,,~~~
\eeq
where $V$ is the volume of the system, $\lambda^{-1}$ characterizes the size of wavepacket,
and $\vec{p}_\rho$ and $\vec{p}_\kappa$ are Jacobi's momenta,
\beq
\vec{p}_\rho = \frac{\, \vec{p}_1 - \vec{p}_2 \,}{\sqrt{2} }\,,
~~~~~~
\vec{p}_\kappa = \frac{\, \vec{p}_1 + \vec{p}_2 -2\vec{p}_3\,}{\sqrt{6} }\,.
~~~~~~
\eeq
The amplitude is large when $\vec{p}_1 \sim \vec{p}_2 \sim \vec{p}_3 \sim \vec{P}/3$. Thus if a baryon has the momentum $\vec{K}\sim \vec{0}$, the quarks inside of it occupy the phase space at low momenta. At low baryon density, low momentum states of quarks are only partially filled, so baryons do not feel the Pauli blocking effect at quark level. As the baryon Fermi sea develops, however, the quark states at low momenta are largely occupied, and a baryon having those states as a part is obstructed; to introduce another baryon into the system, we have to keep the baryon's momentum large enough to avoid conflicts with the quark Pauli blocking.
%

To quantitatively understand whether the quark Pauli blocking becomes significant at density relevant for neutron stars, more details must be worked out by including all quantum numbers (color, spin, flavor) and interactions for baryons. But the above arguments at least illustrate that the hyperons in microscopic descriptions do not emerge as easily as in purely hadronic models. 
%

Within purely hadronic descriptions, we expect that the aforementioned quark Pauli blocking effects can be mimicked by the effective repulsive forces among baryons. This might justify phenomenological introduction of repulsive $YN$ forces. But even so, such treatments work only partially because our ability to handle many-body forces is limited. In contrast, the arguments at quark level have advantages at large density where a hadronic description begins to reveal its weakness.
%

\subsection{On the nature of gluon dynamics}
\label{sec:gluons}

So far our model descriptions have not manifestly taken into account the gluonic degrees of freedom. Some of the gluons may be integrated out into the parameters such as $G_s$, $G_V$, and $H$ in the NJL model. Those parameters are in principle density dependent and should become smaller at larger density as the gluons become weakly coupled due to medium screening effects. In this respect, the impact of the $2M_\odot$ constraint is substantial, since it requires the NJL parameters to be as large as $G_s$ in vacuum. As we saw in the last section, $G_V/G_s>0.5$ and $H/G_s >1.0$ seems necessary for the region of $n_B \sim 5-10n_s$ that corresponds to $\mu\sim 0.5-0.7\,{\rm GeV}$. This suggests that at density relevant to neutron stars the gluon sector remains strongly non-perturbative as in the QCD vacuum\footnote{This point of view was more manifestly examined in Ref.\cite{Fukushima:2015bda}, where the authors discussed how the medium dependent $G_V$ should behave as a function of $\mu$. }.
%

Actually this picture is consistent with another implicit assumption in our model treatments on the QCD vacuum energy. 
%
  
It has been argued that the full QCD vacuum has smaller energy than the perturbative vacuum\footnote{The terminology such as perturbative and non-perturbative vacua might sound awkward as mentioned in Ref.\cite{Shifman:1998rb}, but we will keep using it as long as it does not affect our arguments substantialy.}. In the NJL model, a part of vacuum energy has been automatically taken into account through the Dirac sea contributions associated with the chiral symmetry breaking, ($m_q$: current quark mass, $M_q$: constituent quark mass)
\beq
B_q \equiv \Omega_q(m_q) - \Omega_q (M_q) \sim \lqcd^4 > 0\,,
\eeq
which we call the quark bag constant. On the other hand, this is not the full contribution to the vacuum energy, since non-confining models like NJL do not take into account the long range components of gluons; those gluons, upon integration, would leave non-local interactions in the effective models for quarks, but such interactions are not present in the NJL model\footnote{This two scale argument is analogous to those of Manohar and Georgi \cite{Manohar:1983md}, and Weinberg \cite{Weinberg:2010bq}.}. Therefore besides the Dirac sea contributions, there should be additional energy reduction in the QCD vacuum that is purely caused by the gluons in the infrared. We call it the gluonic bag constant, $B_g>0$, which are related to the gluon condensate. These positive energy contributions should appear whenever the non-perturbative matter changes into perturbative matter; when applying perturbative calculations we must add the positive constant term (bag constant) to the perturbative energy density because such calculations perform an expansion around the trivial vacuum, not around the true QCD vacuum (whose energy density is renormalized to zero at $\mu=T=0$).

Now imagine that the gluons get screened at $n_B\sim 2-10n_s$, and gluons become weakly coupled. Then the size of the gluon condensate is reduced and the gluonic bag constant must be added to (subtracted from) our NJL energy (pressure). But the addition of such constant cannot be as large as $\sim \lqcd^4$, because it would substantially soften the equations of state in which the NJL pressure is only about $0.2-0.5\,{\rm GeV/fm^3} \sim 2\lqcd^4$ at $n_B\sim 5n_s$. Clearly it is no longer possible to pass the $2M_\odot$ constraint. Strictly speaking, even in such situation one can still achieve the necessary stiffness by taking $G_V$ and $H$ much larger than the values we took, but the appearance of the large bag constant and substantially large NJL parameters at finite density seem to be incompatible assumptions. Rather, it is much more consistent to take $G_V$ and $H$ as the values we took, and then also keep the gluonic bag constant to the vacuum value which is much smaller than $\sim \lqcd^4$.

In this view on the medium NJL couplings, our NJL parameters ($G_s, K, G_V, H$) should decrease when the pressure curves approach the pQCD domain. Otherwise the NJL pressure undershoots the pQCD results and the speed of sound is larger. This requirement is again consistent with the idea that the NJL couplings originate from non-perturbative gluons and at very large density should disappear together with such gluons.

Finally we comment on the estimate on $G_V$ which was briefly mentioned in Sec.\ref{sec:vectorint}. In Ref.\cite{Steinheimer:2010sp}, the authors performed the PNJL studies for quark number susceptibilities at $T>T_c$, and showed that those quantities are suppressed too much in the presence of $G_V$. On the other hand, they explained the flatness of the chiral restoration curvature using hadronic models with repulsive density-density interactions for $T<T_c$. But it should be able to express the hadron-hadron interactions in terms of quark degrees of freedom, so we will not distinguish hadron-hadron and quark-quark interactions rigidly\footnote{For instance, the channel dependence of the nucleon-nucleon interactions can be understood by just summing up quark exchange contributions. They can be organized as $1/\Nc^2\sim 1/10$ expansion. See Refs.\cite{Kaplan:1996rk,Hidaka:2010ph} and the references therein. }; in the latter description the repulsive density-density interaction at $T<T_c$ are effectively mimicked by large $G_V$. Thus as explained in Ref. \cite{Ferroni:2010xf}, our interpretation is that $G_V$ is large at $T<T_c$ and small at $T>T_c$. Then we next notice that one of the major characteristic difference between $T< T_c$ and $T>T_c$ is the gluon dynamics. Since we get the NJL parameters after integrating out gluons, the changes of the NJL parameters such as $G_V$ around $T\sim T_c$ are rather natural. We also point out that the changes of the NJL parameters driven by gluons should be universal for all different flavors, and such picture seems to be consistent with the lattice studies on the off-diagonal flavor correlators \cite{Bazavov:2013dta}.

\section{Summary}
\label{sec:summary}

In this paper we discuss the 3-window construction of QCD equations of state, and explain why this construction can express the physics that are largely unexplored in the conventional hybrid or self-bound quark matter equations of state. In the 3-window picture, we have purely hadronic matter, percolated quark matter, and matter intermediate between these two. Such matter is presumably expressed as either hadronic matter about to percolate or quark matter about to be confined. The range of such matter is $n_B\sim  2-5n_s$ or $\mu \sim 0.35-0.6\,{\rm GeV}$.

We examine the significance of the $2M_\odot$ constraint which rules out soft high density equations of state. The constraint is particularly severe when the purely hadronic equations of state are soft: in general it is more difficult to construct equations of state that are soft at low density and stiff at high density, because the combination of them tends to violate the causality constraint. But according to recent results from many-body calculations \cite{Gandolfi:2011xu} and implications of rather small neutron star radii \cite{Steiner:2010fz,Steiner:2012xt,Ozel:2010fw,Ozel:2015fia,Guillot:2014lla,Heinke:2014xaa}, it seems necessary to construct such soft-stiff equations of state. We have explained that the 3-window construction has more chances to provide the soft-stiff equations of state than the conventional hybrid one with the first order phase transition, although the final form of the equations of state can be similar if the strength of the phase transition is very weak.

The 3-window picture also changes the guidelines for practical modeling. Within the NJL model, we found that $G_V>0.5G_s$ and $H>G_s$ are necessary to pass the $2M_\odot$ constraint with the physical speed of sound, while quark model pressure apparently exceeds the hadronic pressure at low $\mu$. In contrast to the conventional hybrid constructions, we accept such behaviors by expecting that at low density the confining effects largely suppress the artificial excess of pressure in the non-confining models. With such expectation, we even think such excess pressure to be quite natural; if the pressure of non-confining models has less pressure than the hadronic one, we must imagine that the confining effects enhance the pressure to approach the full result --- but this is unlikely.

The values of $G_V$ and $H$ used for the percolated quark matter at $n_B\gtrsim 5n_s$ are as large as $G_s$ in the vacuum. Since these NJL parameters should be obtained after integrating out gluons, the large NJL couplings in the percolated domain indicate that the nature of gluons are not changed appreciably by the medium effects to $n_B\sim 10n_s$. This implies that at $n_B \sim 5-10 n_s$ we have the percolated quark matter with non-perturbative gluons which would try to keep the system locally color singlet. The useful regime to study such matter is large $\Nc$, as discussed by McLerran and Pisarski \cite{McLerran:2007qj}.
\\

{\bf Acknowledgement}--- The author acknowledges G. Baym, P.D. Powell, and Y. Song with whom many arguments in this article have been developed. He also appreciates the hospitality of Nuclear theory group at Los Alamos Laboratory, Jilin University, IHEP of CAS, and CCNU where a part of this work was done. He thanks K. Fukushima, J. Stone, and N. Stone for valuable discussions. This work is supported by NSF Grants PHY09-69790 and PHY13-05891.


%
%

\end{document}